\documentclass{elsartL}
\usepackage{graphicx}
\usepackage{amssymb}
\usepackage{amsmath}
\usepackage{physics}
\usepackage{mathrsfs}
\usepackage{mathtools}
\usepackage{appendix}
\newcommand{\avg}[1]{\left< #1 \right>} 

\begin{document}

\begin{frontmatter}
\title{COVID-19: The extraction of the effective reproduction number from the time series of new cases}
\author{Evangelos Matsinos}

\begin{abstract}
Addressed in this work is the performance of five popular algorithms, which aim at assessing the dissemination dynamics of the COVID-19 disease on the basis of the time series of new confirmed cases. The tests are based 
on simulated data, generated by means of a deterministic compartmental epidemiological model \cite{Matsinos2020a}, adapted herein to also include the possibility of the loss of immunity by the group of the recovered (or 
vaccinated) subjects. Assuming a simple temporal dependence of the effective reproduction number (the exact details are of no relevance as far as the conclusions of this work are concerned), time series of new cases were 
generated in a time domain of nearly one year for the five top-ranking countries in the cumulative number of infections by January 1, 2021. These countries are (in descending order of infections): the United States of 
America, India, Brazil, Russia, and the United Kingdom. The processing of each simulated time series led to the establishment of relations between the input (actual) and the reconstructed values of the effective 
reproduction number for each country and algorithm, separately; this work argues that all five algorithms underestimate the effective reproduction number when the latter exceeds the critical value of $1$. The five 
algorithms were subsequently applied to the real-life time series of new cases for the aforementioned five countries, which also span a temporal interval of nearly one year; corrected values of the effective reproduction 
number are obtained for these countries in 2020.\\
\end{abstract}
\end{frontmatter}

\section{\label{sec:Introduction}Introduction}

One year after the World Health Organisation (WHO) declared a pandemia, the `Severe Acute Respiratory Syndrome Coronavirus 2' pathogen (SARS-CoV-2), the etiological agent of the COVID-19 disease, has infected over $2~\%$ 
of the world population thus far, and has claimed well over three million lives. The impact on the economy is felt everywhere, while many countries - in particular in the developing world - struggle to avoid bankruptcy. 
To harness the alarming dissemination of the disease, mitigation measures were implemented in most countries over the course of the past year, enforcing a range of restrictions on the daily activities of the population, 
from moderate ones to curfews and general lockdowns. In parallel with such measures, significant resources were and are being invested in developing, testing, and distributing effective vaccines. By the time this work 
came into completion, the United States Food and Drug Administration (FDA) had formally approved three vaccines \cite{fda}, the Pfizer-BioNTech vaccine by Pfizer Inc.~and BioNTech, the Moderna vaccine by ModernaTX Inc., 
and the Janssen vaccine by Janssen Pharmaceuticals of Johnson \& Johnson Inc. In addition, the European Medicines Agency (EMA), the equivalent of the FDA within the European Union, had approved the Oxford-AstraZeneca 
(now named Vaxzevria) vaccine \cite{ema}, developed by Oxford University and AstraZeneca PLC. The approval of several other vaccines is currently under way, at a time when the support for waiving international patent 
protections for the COVID-19 vaccines is gaining momentum. The general expectation is that these vaccines will reinforce our defences and thus play a decisive role in the global effort to bring the pandemia under control.

Mass-immunisation programmes are in progress with priority given to the so-called high-risk groups, i.e., to the elderly, to the subjects with weakened immune systems (due to underlying medical conditions), and to the 
employees in the healthcare sector. As of May 3, 2021, Israel was leading the vaccination race, with over $62~\%$ of all Israelis already vaccinated at least once \cite{ProgressVaccination}, followed by the United 
Kingdom ($\approx 51~\%$) and the United States of America ($\approx 44~\%$). In most other countries however, the mass-immunisation process appears to be very slow, if not sluggish.

Although not all answers to all questions are known, there can be no doubt that the broad availability of vaccines, in particular for the high-risk groups, is expected to, first, relieve the pressure on the healthcare 
systems and, in the long run, put a halt to the dissemination of the COVID-19 disease. A number of controversial issues are currently being settled regarding the effectiveness of some of the vaccines, in particular 
in terms of the age and of the health status of the vaccinees. Although sporadic cases have surfaced, reporting side effects following vaccination (e.g., thrombotic complications), it must be remembered that no vaccination 
is $100~\%$ free of weak, moderate, or (rarely) severe side effects~\footnote{The decision to proceed or not proceed with a mass-immunisation programme involves a trade-off between the following two risks: i) of 
administering the vaccine to those of the subjects who are not expected to develop life-threatening reactions, and accepting the risk of the development of such reactions, say, at the $p_i$ probability level and ii) of 
taking no action and therefore accepting the disease-induced fatality rate $p_f$. Provided that $p_i \ll p_f$, the risk entailed by mass immunisation is deemed acceptable.}. Finally, it is unclear for how long the 
currently available vaccines may provide immunity, in particular in conjunction with the frequent emergence of virus variants. At this moment, it cannot be excluded that the mass-immunisation process might have to be 
repeated annually, in a way similar to the routine vaccinations against seasonal influenza. Last but not least, the efforts towards the development of more effective vaccines, perhaps variant-specific ones, will 
undoubtedly continue.

During the dissemination of infectious diseases, the extraction of reliable estimates for the effective reproduction number, i.e., for the time-dependent (or instantaneous) reproduction number $R(t)$, serves a number of 
useful purposes.
\begin{itemize}
\item It summarises the `current' situation regarding the dissemination dynamics of the disease.
\item It provides a means for the assessment of the effectiveness of (any) existing mitigation measures, as well as of the possibility of undertaking additional, more stringent ones.
\item It provides a means for the assessment of the effectiveness of the mass immunisation.
\item It enables predictions as to when the herd-immunity level could be reached.
\end{itemize}
The quantity $R(0)$, the quantity associated with the rapidity of the dissemination of a disease in case of i) an uncontaminated sample of susceptible subjects and ii) absence of any mitigation measures, is known in 
Epidemiology as the ``basic reproduction number'' $R_0$. During the course of the dissemination of an infectious disease, the reproduction number generally decreases with time as a result of two effects, namely of the 
depletion of the number of the susceptible subjects (assuming recoveries with immunity) and, more importantly, of the imposition of mitigation measures aiming at curbing exposure.

Several algorithms have been put forward to extract estimates for the effective reproduction number from the time series of new cases of confirmed infections (new cases, for short). The objective in this work is to test 
some of these methods with data obtained under controlled conditions, on the basis of a recently introduced model tailored to the characteristics of the COVID-19 decease \cite{Matsinos2020a}. Input in the generation of 
each time series of new cases is the dependence of the effective reproduction number on time. The idea towards performing such a test occurred to me as, for a large part of 2020, presented in the media were reports/plots 
of decreasing (and dropping below the critical value of $1$) values of the effective reproduction number (therefore marking the beginning of the end of the dissemination of the disease), just before the second wave of 
the pandemia started setting in. It might be that this conundrum could be explained as the result of undue optimism and consequent wilful relaxation of the mitigation measures; of unintentional relaxation of the 
mitigation measures (e.g., due to habituation); or of a tendency of the established algorithms to yield underestimates of the effective reproduction number in case of this disease, thus portraying a more optimistic state 
of affairs than it actually was. This study examines the last option; understanding the origin of the aforementioned discrepancy is expected to improve the planning, as well as the tactics employed in the war against 
this virus.

In the last part of the study, the analysis of real-life data will be pursued, making use of the time series of new cases corresponding to the five top-ranking countries in terms of the cumulative number of infections 
by January 1, 2021. These countries are (in descending order of infections): the United States of America (USA), India (IND), Brazil (BRA), Russia (RUS), and the United Kingdom (GBR). The real-life data have been obtained 
from the European Centre for Disease Prevention and Control \cite{ECDC}.

The structure of this paper is as follows. The model, used in the generation of the time series of new cases, is detailed in Section \ref{sec:Model}: Sections \ref{sec:Model_Definitions}, \ref{sec:Model_Parameters}, and 
\ref{sec:Model_ODEs} introduce the groups of this model, the model parameters, and the mathematics involved in the time evolution of the populations of the model groups, respectively. Section \ref{sec:EffTr} provides a 
short description of the established algorithms aiming at extracting the effective reproduction number from the time series of new cases. In the first part of the subsequent section, Section \ref{sec:Results}, these 
methods are applied to the time series of new cases, generated using the central parameter values of Table \ref{tab:DiseaseRelatedParameters} and the time evolution of the populations of the model groups of Section 
\ref{sec:Model_ODEs}; in the second part of that section, the corrections, obtained in the first part, are applied to the aforementioned five real-life time series of new cases. The last section of this study, Section 
\ref{sec:Conclusions}, provides a concise overview of the findings of this work and discusses the implication of these findings in the planning against the COVID-19 disease.

Unless otherwise stated, the unit of time in this work is one day (d).

\section{\label{sec:Model}The model}

\begin{figure}
\begin{center}
\includegraphics [height=10cm,angle=90] {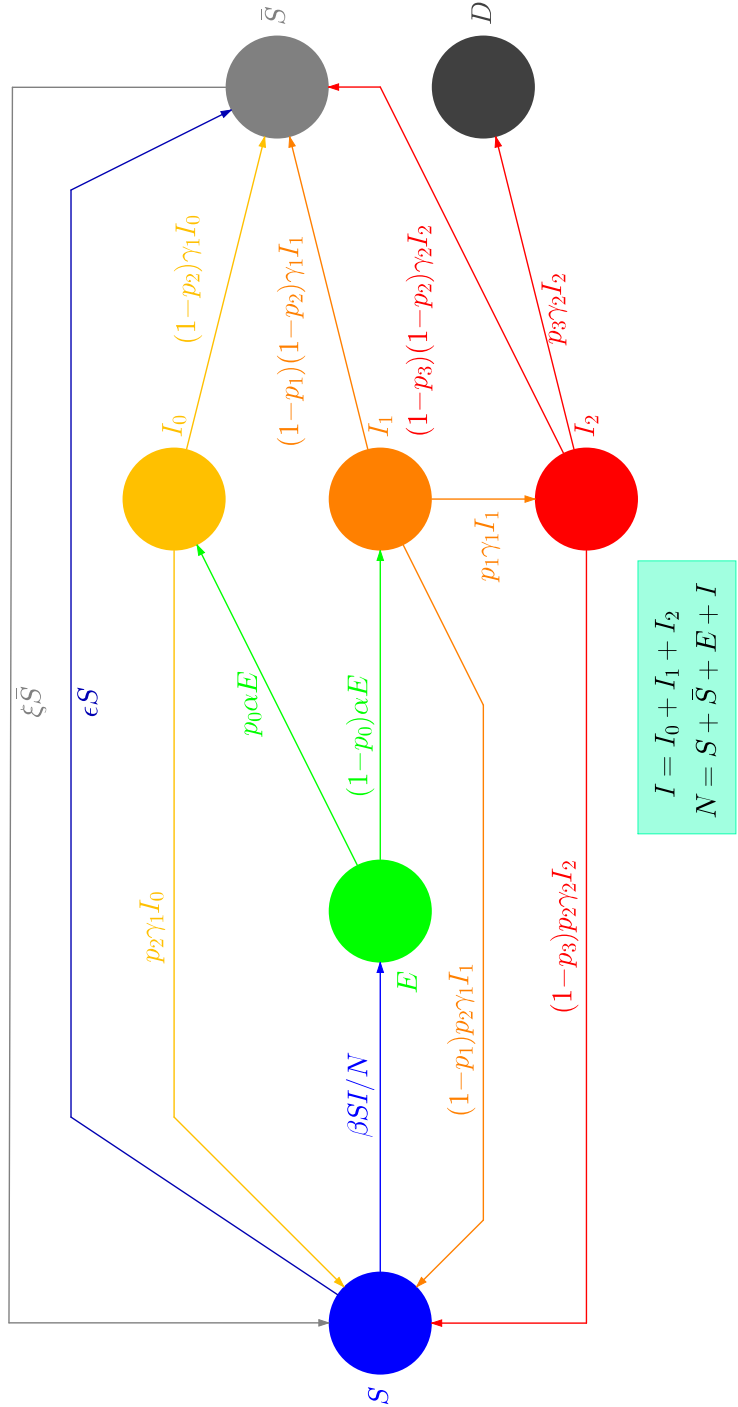}
\caption{\label{fig:CompartmentalModelForCOVID19}The transfer diagram of the deterministic compartmental epidemiological model of this work, essentially introduced in Ref.~\cite{Matsinos2020a}. Included herein is the 
possibility of the loss of immunity by the group of the recovered (or vaccinated) subjects $\bar{S}$ (contribution shown in grey, equal to $\xi \bar{S}$). Although not explicitly indicated in this diagram, the vital 
dynamics, i.e., the new births and the natural (i.e., not disease-induced) fatalities, are taken into account in the time evolution of the populations of the model groups.}
\vspace{0.35cm}
\end{center}
\end{figure}

A useful review of the efforts to model the dissemination of infectious diseases may be found in Ref.~\cite{Hethcote2000}; the first compartmental models in Epidemiology are approximately one century old. The transfer 
diagram of the model of this work is displayed in Fig.~\ref{fig:CompartmentalModelForCOVID19}. Evidently, used herein is the complete (i.e., including the vital dynamics and the effects of the mass immunisation) model 
of Ref.~\cite{Matsinos2020a} (see Appendix A therein) with one addition which will be addressed in the subsequent section. As the case is with most compartmental epidemiological models, the application of this model may 
involve the human population within a region of interest (i.e., within one country, province, city) or subgroups of that population.

\subsection{\label{sec:Model_Definitions}Definitions}

At a specific (calendar) time $t$, each subject is categorised into one of the following seven groups.
\begin{itemize}
\item The `susceptible' subjects ($S$) are those who can be infected.
\item The `exposed' subjects ($E$) are those who have been infected, but are not yet infective themselves.
\item The `asymptomatic infective' subjects ($I_0$) are those who have been infected, but - whichever the reason - will not develop symptoms (at least in the time domain of the specific study).
\item The `symptomatic phase-I infective' subjects ($I_1$) are infective subjects who are developing weak symptoms.
\item The `symptomatic phase-II infective' subjects ($I_2$) are infective subjects who are developing severe symptoms (necessitating hospitalisation).
\item The `unsusceptible' subjects ($\bar{S}$) are those subjects who are (at time $t$) immune. The reason for the immunity is of no relevance: these subjects might have natural immunity against the disease, 
they might have acquired immunity after vaccination, or they might have acquired immunity after they contracted (and recovered from) the disease.
\item The `deceased' subjects ($D$) are those who did not survive the phase II of the disease, as well as those who died (within the time domain of the study) due to other (i.e., not disease-specific) reasons.
\end{itemize}
Evidently, these seven populations~\footnote{For the sake of brevity, the same letters will be used henceforth, to identify both the groups as well as their sizes (populations). For instance, $S$ will denote both the 
group of the susceptible subjects as well as the (time-dependent) population of that group.} are time-dependent.

The only difference to the model of Ref.~\cite{Matsinos2020a} is that the unsusceptible subjects $\bar{S}$ revert to the group of the susceptible subjects $S$ at a rate $\xi$. At this moment, there is a paucity of studies 
reporting on $\xi$ (which, of course, is hardly surprising, given that this is a novel virus); recent studies \cite{Gudbjartsson2020,Ripperger2020,Rodda2021} have come up with a lower bound on $\xi^{-1}$, demonstrating 
that most subjects, who had recovered from the disease, acquired immunity for \emph{at least} a few months. In congruence with the widespread speculation that the mass-immunisation process will have to be repeated 
periodically, $\xi^{-1}$ will be assumed herein equal to a temporal interval of nine months. Provided that this interval is not much shorter, the time series, entering the tests in this work, will - for all practical 
purposes - be independent of the choice of the value of this parameter. Evidently, the model of Ref.~\cite{Matsinos2020a} may be retrieved by setting $\xi$ to $0$.

The discussion will be facilitated if, prior to entering any details, some additional definitions are given. As far as the early development of infectious diseases is concerned, three are the important moments:
\begin{itemize}
\item the time $t_1$ at which a subject becomes infected,
\item the time $t_2$ at which the subject becomes infective, and
\item the time $t_3$ marking the clinical onset (onset of symptoms) for that subject.
\end{itemize}
(For some diseases, $t_2>t_3$: an infected subject becomes infective after the onset of symptoms. In the case of the COVID-19 disease, there is ample evidence to support frequent pre-symptomatic transmission 
\cite{Ganyani2020,Wei2020,Arons2020,Ren2021,Zhang2021}.) Were the development of a disease identical for all infected subjects, an infector and a related infectee would follow time-shifted, yet identical, timelines: if 
$t_1$, $t_2$, and $t_3$ are the infector's timeline characteristics, those of the infectee would be: $t_1^\prime=t_1+\Delta t$, $t_2^\prime=t_2+\Delta t$, and $t_3^\prime=t_3+\Delta t$, where $t_2-t_1 \leq \Delta t \leq t_2 - t_1 + t_{\rm inf}$; 
the quantity $t_{\rm inf}$ denotes the temporal interval within which the infector remains infective (also known as infectivity interval). However, neither the infection-to-infectiousness ($t_2-t_1$) nor the incubation 
($t_3-t_1$) interval may be thought of as $\delta$-distributed quantities; in case of the COVID-19 disease, broad distributions have been established.

From the aforementioned three time instants, $t_3$ is the easiest to pin down. In case of $t_1$, a temporal interval can frequently be established, representing the most likely time span during which the infection 
occurred; for instance, if the infected subject had visited a high-risk region (in the recent past), the start and the end time instants of that trip provide estimates for the lower and upper bounds of $t_1$. The 
determination of $t_2$ is less straightforward. To describe the characteristics of the dissemination of infectious diseases, Epidemiology makes use of three temporal intervals~\footnote{Concerning the definitions of 
these intervals, there is some confusion in the literature; for instance, the serial and the generation intervals were, until recently, used interchangeably.}:
\begin{itemize}
\item the infection-to-onset-of-symptoms interval (incubation interval) $t_3-t_1$ (defined for each subject),
\item the onset-to-onset interval (serial interval) $t_3^\prime-t_3$ (defined for each infector-infectee pair), and 
\item the infection-to-infection interval (generation interval) $t_1^\prime-t_1$ (defined for each infector-infectee pair).
\end{itemize}
Naively, one would expect that i) these three intervals are positive and ii) the serial and generation intervals are about equal. In case of the COVID-19 disease however, only the incubation and the generation intervals 
are positive; as aforementioned, Refs.~\cite{Ganyani2020,Wei2020,Arons2020,Ren2021,Zhang2021} have reported that the PDF of the serial interval contains substantial negative tails, implying that - on occasion - the 
infectee develops symptoms before the infector does.

Interesting in the context of most standard epidemiological models (as well as of the model of this work) is the infection-to-infectiousness interval $t_2-t_1$. Evidently, none of the aforementioned three (incubation, 
serial, and generation) intervals, which are routinely reported in epidemiological studies, appears to be relevant. In the absence of appropriate data, an acceptable estimate for the infection-to-infectiousness interval 
$t_2-t_1$ could be the generation interval. However, the general paucity of information about this quantity enforces the use of the serial interval in many studies. According to Caicedo-Ochoa and collaborators 
\cite{Caicedo2020}, as well as to Knight and Mishra \cite{Knight2020}, both the choice of the quantity, which should be associated with the development of the disease from the time of a primary to that of a secondary 
infection, as well as the properties of the PDF of that quantity matter in the evaluation of the effective reproduction number.

\subsection{\label{sec:Model_Parameters}The model parameters}

The complete version of the model calls for twelve parameters, and may be applied to the long-term dissemination of infectious diseases similar to the COVID-19 disease. The nine disease-specific model parameters are as 
follows.
\begin{itemize}
\item $\alpha$ is the rate parameter associated with the exponential decrease (excluding vital dynamics) in the population of the exposed $E$, provided that the transmission process ceases. This parameter is related to 
the infection-to-infectiousness interval; in the absence of relevant information, it appears to be more appropriate to associate $\alpha$ with the average generation interval (rather than with the average incubation or 
serial interval).
\item $\beta$ is the infection rate. Being the product of the average number of contacts per person per unit time and the probability of virus transmission in a contact between an infective and a susceptible subject, 
this parameter may be thought of as the propellant for the dissemination of the disease.
\item The parameters $\gamma_{1,2}$ reflect the rates at which the populations $I_{1,2}$, respectively, would diminish (excluding vital dynamics), were they not replenished by new symptomatic infective subjects. The 
decrease in the $I_1$ population is due either to the recovery of the subject or to the development of severe symptoms (in which case, the subject enters the phase II of the COVID-19 disease and the $I_2$ population). 
The decrease in the $I_2$ population is due either to the subject's recovery or to the subject's (disease-induced) death.
\item $p_0$ is the fraction of the exposed who become asymptomatic subjects ($E \to I_0$). The remaining subjects, departing from group $E$, enter the $I_1$ population.
\item $p_1$ is the fraction of the symptomatic phase-I infective subjects who enter the phase II of the COVID-19 disease ($I_1 \to I_2$).
\item $p_2$ is the fraction of the recoveries without (temporary) immunity.
\item $p_3$ is the fatality rate, defined here as \emph{the fraction of the symptomatic phase-II infective subjects who do not recover} ($I_2 \to D$).
\item $\xi$ is the rate parameter associated with the exponential decrease (excluding vital dynamics) in the population of the unsusceptible subjects, provided that the transmission process ceases ($\bar{S} \to S$).
\end{itemize}

In addition, three model parameters must be fixed for each country separately: two of these parameters relate to the inclusion of the vital dynamics, whereas the third accounts for the effects of the mass immunisation.
\begin{itemize}
\item The annual birth rate will be denoted by $\lambda$.
\item The annual mortality rate will be denoted by $\mu$. Of course, to avoid double counting, the disease-induced mortality rate must be removed from the reported $\mu$ estimates for 2020. This correction, which is 
equal on average to about $0.24 \cdot 10^{-3}$ per capita, varies widely among the countries; for instance, it is equal to about $0.11 \cdot 10^{-3}$ per capita for IND, about ten times as high for the USA and GBR.
\item The mass-immunisation rate will be denoted by $\epsilon$.
\end{itemize}

The model, proposed in Ref.~\cite{Matsinos2020a} and slightly modified herein to account for the loss of immunity by the group of the recovered (or vaccinated) subjects, reduces to the standard epidemiological model 
SEIR (the group $R$ of the SEIR model is obviously identified with the model group $\bar{S}$) for a suitable choice of its parameters (and of the initial conditions for the populations); this is achieved by setting 
$p_0=p_1=p_2=\xi=\epsilon=0$, by identifying $\gamma_1$ with the (one) recovery rate $\gamma$, and by setting the initial populations $I_0(0)$, $I_2(0)$, and $D(0)$ to $0$. To retrieve the SIR model, one must remove 
the group $E$ from the SEIR model (thus enable the direct transition $S \to I$). Finally, the model SIS may be retrieved from the model of this work after following the aforementioned steps resulting in the SIR model, 
but setting $p_2$ to $1$; of course, $\bar{S}(0)=0$ in the last case. In case of the SIR and SEIR models, the group $\bar{S}$ may be defined in such a way as to contain the group of the deceased $D$.

Estimates for the nine disease-specific parameters of this work are listed in Table \ref{tab:DiseaseRelatedParameters}. The values of the two country-related parameters (for the five countries mentioned at the end of 
Section \ref{sec:Introduction}), as well as the population in each case (which enters the initial conditions in the determination of the time evolution of the populations of the model groups) are given in Table 
\ref{tab:CountryRelatedParameters}. As the real-life data had been acquired at times when no vaccines were available, the parameter $\epsilon$ will be fixed to $0$ henceforth.

Regarding the dynamics of the model, a number of remarks are due.
\begin{itemize}
\item All newborns are assumed to enter the group of the susceptible subjects (that is, there is no maternal immunity or association with their mothers' group).
\item From the point of view of the infectiousness and the recovery rate, the asymptomatic infective subjects are treated as if they were symptomatic phase-I infective subjects; therefore, unlike in other studies, it 
will not be assumed herein that the $I_0$ subjects are less infective than the subjects of the other two infective groups $I_{1,2}$.
\item The infectivity interval is assumed to be identical to the entire interval relating to the recovery (or until death for those of the phase-II patients who do not recover).
\item The recoveries without (temporary) immunity involve the same probability ($p_2$) regardless of the type of the infective subject (i.e., whether that subject originally belonged to the $I_0$, $I_1$, or $I_2$ group). 
Some information about the parameter $p_2$ may be extracted from Ref.~\cite{Gudbjartsson2020}: it was found therein that $1\,107$ out of the $1\,215$ subjects, who had recovered from the disease and were submitted to 
serological tests, had developed increasing amounts of antibodies in the first two months following the diagnosis. As a result, Ref.~\cite{Gudbjartsson2020} suggests that $p_2=8.89(82)~\%$.
\item Two additional parameters were used in Ref.~\cite{Matsinos2020a} to model the temporal dependence of the effective reproduction number. As the effective reproduction number is part of the \emph{input} in this work, 
the functional dependence of this quantity on time $t$ is analytically known.
\end{itemize}

\begin{table}
{\bf \caption{\label{tab:DiseaseRelatedParameters}}}Average values and $1\sigma$ uncertainties of the nine disease-specific model parameters. The uncertainty of the quantity $\alpha$ should be considered approximate; it 
was obtained in Ref.~\cite{Matsinos2020a} from an analysis of reported results for the serial interval. It is unlikely that the various virus variants share the same parameter values; in fact, it has been argued that 
some virus variants are more aggressive (higher infection rate) or more lethal (higher fatality rate) than others, etc. The parameter values in this table were extracted from results obtained during the early course of 
the COVID-19 disease. To be suitable for applications involving the recent (e.g., UK, South-African, Brazilian, Indian) virus variants, it is very likely that these values require revision.
\vspace{0.2cm}
\begin{center}
\begin{tabular}{|c|c|c|c|}
\hline
Parameter & Description & Value & Source\\
\hline
\hline
$\alpha$ & The inverse of the average & $0.251(13)$ d$^{-1}$ & \cite{Knight2021}\\
 & generation interval & & \\
$\beta$ & The infection rate & Variable, see Section \ref{sec:Model_Final} & \\
$\gamma_1$ & The inverse of the average & $0.201(36)$ d$^{-1}$ & \cite{Matsinos2020a}\\
 & recovery interval for phase I & & \\
$\gamma_2$ & The inverse of the average & $0.0658(66)$ d$^{-1}$ & \cite{Matsinos2020a}\\
 & recovery interval for phase II & & \\
$p_0$ & The fraction $E \to I_0$ & $17.0(1.0)~\%$ & \cite{Matsinos2020a}\\
$p_1$ & The fraction $I_1 \to I_2$ & $18.52(18)~\%$ & \cite{Matsinos2020a}\\
$p_2$ & The fraction of recoveries & $8.89(82)~\%$ & \cite{Gudbjartsson2020}\\
 & without (temporary) immunity & & \\
$p_3$ & The fraction $I_2 \to D$ & $12.32(40)~\%$ & \cite{Matsinos2020a}\\
$\xi$ & The inverse of the average & $0.0037(12)$ d$^{-1}$ & Approximate estimate on\\
 & immunity-loss interval & & the basis of Refs.~\cite{Gudbjartsson2020,Ripperger2020,Rodda2021}\\
\hline
\end{tabular}
\end{center}
\vspace{0.5cm}
\end{table}

\begin{table}
{\bf \caption{\label{tab:CountryRelatedParameters}}}Values of the two country-dependent model parameters, i.e., of the annual birth rate $\lambda$ and of the annual mortality rate $\mu$, both normalised to $1\,000$ 
subjects. The birth-rate data have been obtained from the Population Reference Bureau \cite{PRB} and correspond to the year 2020. The mortality-rate data have been obtained from the World Factbook of the Central 
Intelligence Agency \cite{CIA} and also correspond to the year 2020; subtracted from the values of Ref.~\cite{CIA} (and explicitly given in the table) are the estimates for the mortality rate relating to the COVID-19 
disease, obtained from the cumulative number of disease-induced fatalities in each country by January 1, 2021 \cite{worldometers}. The population of each country has been fixed from the `World Population Review' 
\cite{WP} and corresponds to their column `2020 Population' (numbers relating to July 1, 2020, copied on February 27, 2021).
\vspace{0.2cm}
\begin{center}
\begin{tabular}{|c|c|c|c|}
\hline
Parameter & Birth rate $\lambda$ & Mortality rate $\mu$ & Total population\\
\hline
\hline
USA & $12$ & $8.30-1.09$ & 331\,002\,651\\
IND & $20$ & $7.30-0.11$ & 1\,380\,004\,385\\
BRA & $14$ & $6.90-0.92$ & 212\,559\,417\\
RUS & $11$ & $13.40-0.39$ & 145\,934\,462\\
GBR & $11$ & $9.50-1.08$ & 67\,886\,011\\
\hline
\end{tabular}
\end{center}
\vspace{0.5cm}
\end{table}

\subsection{\label{sec:Model_ODEs}Time evolution}

The transfer diagram of Fig.~\ref{fig:CompartmentalModelForCOVID19} leads to the following system of first-order ordinary differential equations (ODEs).
\begin{align} \label{eq:EQ0001}
\dot{S} &= -\beta S I/N + p_2 \gamma_1 I_0 + (1 - p_1) p_2 \gamma_1 I_1 + (1 - p_3) p_2 \gamma_2 I_2 - (\epsilon + \mu) S + \xi \bar{S} + \lambda N \, \, \, , \nonumber\\
\dot{\bar{S}} &= (1 - p_2) \gamma_1 I_0 + (1 - p_1) (1 - p_2) \gamma_1 I_1 + (1 - p_3) (1 - p_2) \gamma_2 I_2 + \epsilon S - (\xi + \mu) \bar{S} \, \, \, , \nonumber\\
\dot{E} &= \beta S I/N - (\alpha + \mu) E \, \, \, , \nonumber\\
\dot{I_0} &= p_0 \alpha E - (\gamma_1 + \mu) I_0 \, \, \, , \nonumber\\
\dot{I_1} &= (1 - p_0) \alpha E - (\gamma_1 + \mu) I_1 \, \, \, , \nonumber\\
\dot{I_2} &= p_1 \gamma_1 I_1 - (\gamma_2 + \mu) I_2 \, \, \, , \text{and} \nonumber\\
\dot{D} &= p_3 \gamma_2 I_2 + \mu N \, \, \, ,
\end{align}
where the dots denote derivatives with respect to time, $I=I_0+I_1+I_2$ is the total number of infective subjects, and $N=S+\bar{S}+E+I$ is the total number of living subjects at a specific time. For fixed parameter values 
and for a specific temporal dependence of the effective reproduction number, the time evolution of the seven populations may be obtained via standard methods (see the beginning of Section \ref{sec:TestResults}). It goes 
without saying that the prerequisite to the validity of the extracted solution for these populations (in terms of compatibility with the observations) is the homogeneous mixing of the populations of the susceptible and 
infective subjects.

The interest in this work lies in the analysis of the time series of new cases, i.e., of the sequence of numbers of reported infections between an instant $t$ and another instant $t+\tau$, where $\tau>0$. As the $I_0$ 
subjects have no reasons to suspect that they are infected, it is unlikely that they will subject themselves to testing. The assumption is that all $I_1$ subjects will submit themselves to testing regardless of the 
severity of their symptoms, and that all these tests will be $100~\%$ reliable, confirming infection. As a result, the generated time series of new cases will be associated with the integral
\begin{equation} \label{eq:EQ0002}
\mathscr{I} (t; \tau) \coloneqq (1 - p_0) \alpha \int_{t}^{t+\tau} E(t^\prime) dt^\prime \, \, \, ,
\end{equation}
whereas those of new (disease-induced) fatalities (not used in this work, but given for the sake of completeness) with the integral
\begin{equation} \label{eq:EQ0003}
\mathscr{F} (t; \tau) \coloneqq p_3 \gamma_2 \int_{t}^{t+\tau} I_2(t^\prime) dt^\prime \, \, \, .
\end{equation}

\subsection{\label{sec:Model_Final}Final touches}

The following simple parameterisation of the temporal dependence of the effective reproduction number was implemented:
\begin{equation} \label{eq:EQ0004}
R(x=t/t_{\rm tot}) = P - (P-L) x \, \, \, ,
\end{equation}
where $t_{\rm tot}$ is the temporal interval within which the solution of the system of ODEs of Eqs.~(\ref{eq:EQ0001}) is sought. Evidently, the quantities $P$ and $L$ represent $R(0)$ and $R(1)$, respectively. The 
parameter $L$ was set equal to $0.90$ in all cases, whereas the parameter $P$ was varied for each country separately, in order that an approximate agreement be obtained for the cumulative number of infections between 
the solution of the system of ODEs of Eqs.~(\ref{eq:EQ0001}) and the information contained in the real-life data; it must be emphasised that, provided that the two numbers remain reasonably close (e.g., within a few 
percent of one another), the level of agreement between them is inessential for the purposes of this work, as are the details of the parameterisation of the temporal dependence of the effective reproduction number.

The basic reproduction number $R_0$ is defined as the ratio between the (nominal) infection and the recovery rates, $\beta/\gamma$. This definition brings up the question of how $R_0$ should be defined in the presence 
of two $\gamma$ values, e.g., pertaining to different stages of infectious diseases, as the case is herein. Without doubt, the appropriate definition of the recovery rate should take account of both phases of the COVID-19 
disease, as well as of the probability that a phase-I infective subject enter the phase II of the disease. In Ref.~\cite{Matsinos2020a}, the average recovery interval for phase-I subjects was inferred from the temporal 
interval between the onset of symptoms and hospitalisation for the phase-II subjects. In essence, the hospitalisation was interpreted in Ref.~\cite{Matsinos2020a} as evidence that the phase I of the COVID-19 disease was 
completed without recovery. On the basis of that assumption, the recovery interval for phase I was found equal to $T_1=4.98(90)$ d. On the other hand, it appears that the development of the COVID-19 disease for patients 
with severe symptoms takes more time: on average, these subjects recover (or succumb to the disease) $T=20.2(1.2)$ d after the onset of symptoms. One may therefore associate the average recovery rate for the phase 
II of the COVID-19 disease with the difference of the two recovery intervals: $T_2=T-T_1=15.2(1.5)$ d. An estimate for the effective recovery interval may be obtained via the expression:
\begin{equation} \label{eq:EQ0007}
\avg{T}=(1 - p_1 + p_0 p_1) T_1 + (1 - p_0) p_1 T = T_1 + (1 - p_0) p_1 T_2 = 7.32(79) \text{d} \, \, \, .
\end{equation}
One thus obtains $\avg{\gamma}=1/\avg{T}=0.137(15)$ d$^{-1}$ and $R_0 \coloneqq \beta / \avg{\gamma}$. It follows that an effective infection rate $\beta (t)$, reflecting the impact of the mitigation measures on the 
dissemination of the COVID-19 disease, may be obtained as the product of the effective reproduction number $R(t)$ and $\avg{\gamma}$.

Of course, any procedure, which is put forward as a means to determine the effective reproduction number on the exclusive basis of a time series of new cases, raises one question: how can an observable, which relates 
only to the onset of symptoms of a disease (e.g., involving only the parameters $\alpha$ and $\beta$ in the model of this work), yield any information about a quantity which also includes characteristics of the disease 
(namely the parameter $\avg{\gamma}$ in the model of this work) which can only be assessed and evaluated \emph{after} the onset of symptoms? The answer to this question is that the characteristics of a disease after 
the onset of symptoms are indirectly encoded in the time series of new cases, in that the infectivity interval is associated with the recovery interval(s) of that disease. Many works identify the effective reproduction 
number with the effective transmissibility, which is formally defined as the average number of secondary infections per primary infection, i.e., the number of subjects who become infected by an infective subject on 
average \cite{Lipsitch2003,Cauchemez2006}: according to this definition, the effective transmissibility is equal to the product of the effective infection rate $\beta(t)$ and the average infectivity interval. On the 
other hand, the effective reproduction number, as it was originally defined in epidemiological modelling obeys: $R(t) = \beta (t) / \gamma$ (or $R(t) = \beta (t) / \avg{\gamma}$ in the context of this work).

If the average infectivity interval $\avg{t_{\rm inf}}$ is identified with the average recovery interval $\avg{T}$ of Eq.~(\ref{eq:EQ0007}), then the effective reproduction number and the effective transmissibility are 
indistinguishable measures of the dissemination dynamics of the disease. On the other hand, if $\avg{t_{\rm inf}}$ turns out to be shorter than $\avg{T}$ (e.g., as a result of the isolation of the majority of the 
infective subjects), the effective transmissibility, as estimated by the various algorithms, will \emph{inevitably} be an underestimate of the effective reproduction number. The time series of new cases are bound to 
contain the effects of the isolation of the infective subjects, in that the majority of these subjects will not be able to generate new infections, at least as easily as they would have, had they been able to remain in 
ordinary contact with the population of the susceptible subjects. At the same time, the pressure on the healthcare systems (which only the effective reproduction number quantifies) would not be relieved in any way, for 
the simple reason that, isolated as they might be, the infective subjects would still need medical care, and (on occasion) intensive-care and ventilation beds. To summarise, though the effective transmissibility might 
be a practicable indicator of the dissemination dynamics in the long run, it does not seem to be helpful when it comes down to planning for the short-term demands on the healthcare facilities.

\section{\label{sec:EffTr}An overview of methods for extracting the effective transmissibility from the time series of new cases}

This section provides a short overview of some popular algorithms used in the determination of the effective transmissibility from the time series of new cases. Although one may find most details in the original papers, 
this repetition is inevitable, not only due to didactical reasons, but also to keep this work largely self-contained.

In 2003, Farrington and Whitaker extracted $R(t)$ for mumps and rubella at two times, before and after vaccines were administered to the population in England and Wales \cite{Farrington2003}. Also in 2003, Lipsitch and 
collaborators made use of epidemiologic data from Singapore and extracted $R_0$ in case of the first Coronavirus (SARS-CoV-1), which had caused another epidemic outbreak in the early years of this millennium 
\cite{Lipsitch2003}. In their paper, the authors also assessed the impact of the mitigation measures on the dissemination of that disease, and made interesting remarks about the asymptomatic subjects (which are equally 
applicable to COVID-19): ``If asymptomatically infected persons become immune to subsequent infection without suffering from SARS, this will ultimately reduce transmission by reducing the susceptible population. However, 
if asymptomatic persons contribute substantially to transmission but are not readily identified as SARS cases, control measures will be hampered because they depend on the ready identification of people who have been 
exposed to potentially infectious cases.''

To suppress the statistical fluctuations in the input (which they recognised as an obstacle to extracting reliable estimates for the epidemiological parameters of infectious diseases), Bettencourt and Ribeiro developed 
in 2008 a Bayesian method \cite{Bettencourt2008} and applied it to the time series of infections~\footnote{It is my understanding that the method of Bettencourt and Ribeiro should use as input the time series of the 
\emph{active} cases.} relevant to the H5N1 influenza (commonly known as `bird flu'), a disease with a high fatality rate which became a major concern for several countries in southeast/east Asia during the mid 2000s.

One year later, Cintr{\'o}n-Arias and collaborators developed a method for the estimation of $R(t)$ (and its assigned uncertainty), featuring fits of a deterministic SIR epidemiological model to the time series of new 
infections \cite{Cintron2009}. They applied their approach to data acquired during the outbreaks of seasonal influenza (H3N2), over eight consecutive years starting in October 1997.

Before entering the analysis methods which will be used in this work to provide estimates for the effective transmissibility, I will give some details about the input, the prime objective in its analysis, and some 
features which these methods share.
\begin{itemize}
\item The input is a discrete-time signal, comprising the time series of new cases, as reported daily by the appointed authorities in the public-health sector in all sovereign countries around the world. The start of 
each day is conventionally set to 00:00:00 UTC. The number of (full) days (elapsed since $t=0$) will be identified by an index $i \in \{ 0, \dots, n \}$ for $n \in \mathbb{Z}^+$, $(n+1) \tau$ being equal to the temporal 
interval $t_{\rm tot}$ within which the solution of the system of ODEs of Eqs.~(\ref{eq:EQ0001}) is sought; in mathematical notation, day $i$ spans the temporal interval $[i \tau,(i+1) \tau)$, where $\tau=86400$ s ($1$ 
d). The number of new cases on day $i$, to be identified with $\mathscr{I} (t=i \tau; \tau)$ of Eq.~(\ref{eq:EQ0002}), will be denoted henceforth as $\mathscr{I}_i$. The input comprises sequences of values 
$\mathscr{I}_0, \dots, \mathscr{I}_{n}$.
\item In the real-life data mentioned in Section \ref{sec:Introduction}, the clock was set to $0$ on December 31, 2019, 00:00:00 UTC; as a result, the first day in the observations (i.e., the one corresponding to the 
day identifier $i=0$) was the last day of 2019. Each array in the real-life data starts with a segment containing several null values. The last reported day in Ref.~\cite{ECDC} was December 14, 2020; therefore, $n=349$ 
in the real-life data.
\item The methods of this section aim at establishing their `best' estimator to extract from the input time series of new cases the effective transmissibility on day $i$, to be denoted henceforth as $R(t_i)$.
\item Three of the methods of this section feature an array of weights $w_j$, for $j= \{ 0, \dots, M-1 \}$ and $M \in \mathbb{Z}^+$; these weights are also known as infectivity weights. Albeit user-defined, hence 
arbitrary, the dimension $M$ of the array of weights may be established from the properties of the assumed PDF of the serial interval in such a way that the temporal interval of $M$ days covers the region of interest 
(i.e., the essential part) of that distribution. (Of course, each element of the array of weights will be obtained as the difference of the cumulative distribution function (CDF) of the serial interval at the two relevant 
endpoints $j$ and $j+1$.) As the array of weights must be truncated in the application, its elements will be renormalised, i.e., they will be redefined in such a way as to finally fulfil the condition: $\sum_{j=0}^{M-1} w_j = 1$. 
The dimension $M$ will be chosen such that the CDF of the serial interval exceed the value $0.99$ at the upper end of the last bin, i.e., of the one corresponding to the last element of the array of weights.
\item There is confusion regarding the infectivity weights, namely whether they better be associated with the serial or with the generation interval. A simple answer to this question may be found in Ref.~\cite{Nishiura2009}: 
in one short sentence, the choice depends on the type of the observable. As the real-life data involve the number of new cases, the time instant of the onset of symptoms is relevant, not the one at which the corresponding 
infection occurred. As a result, it appears that the serial interval is the appropriate physical quantity to be associated with the infectivity weights. From available results, extracted in Ref.~\cite{Matsinos2020a} were 
the values: $\nu=4.35(25)$ d for the average and $\sigma=4.66(21)$ d for the rms (root mean square) of the PDF of the serial interval.
\item The PDF of the serial interval has been modelled in a variety of ways in the literature: normal, lognormal, Weibull, and gamma distributions have been used; of course, only the first of these distributions has a 
negative tail. In this work, the infectivity weights will be obtained on the basis of the Gaussian distribution $N(\nu,\sigma^2)$ in the domain $0 \leq x \leq M$, for the aforementioned values of the parameters $\nu$ 
and $\sigma$.
\item Provided that the PDF of the serial interval is the normal distribution with the aforementioned parameter values, pre-symptomatic transmissions are expected to occur at the level of about $17.5~\%$ of all 
infections. This result is about one-third of the estimates obtained in Refs.~\cite{Ganyani2020,Arons2020,Ren2021}, a weighted average of which would be equal to $47.1(5.2)~\%$; on the other hand, such a large probability 
of pre-symptomatic transmissions is not supported by Ref.~\cite{Wei2020}.
\item There is confusion also regarding (some of) the definitions in the works providing the algorithms. It must be reminded that the input data comprises the numbers of subjects with onset of symptoms on 
specific days; the time instants at which these subjects became infected or became infective themselves are of no relevance.
\end{itemize}

For the remaining part of this section, I will enter some of the details of the algorithms which will be used in the subsequent section in order to provide estimates for the effective transmissibility.

\subsection{\label{sec:WT}The Wallinga-Teunis (WT) method}

The Wallinga-Teunis method \cite{Wallinga2004} features a statistical procedure which aims at assessing the likelihood that two cases $j$ and $i$, appearing in the time series of new infections and respectively 
corresponding to times $t_j$ and $t_i \geq t_j$, be causally connected, i.e., be identified as an infector-infectee pair. They introduced this probability as the ratio
\begin{equation} \label{eq:EQ0008}
p_{ji}=\frac{w(t_i-t_j)}{\sum_{k<i} w(t_i - t_k)}
\end{equation}
and proceeded to define the transmissibility, associated with the infector $j$, as the sum
\begin{equation} \label{eq:EQ0009}
R_j = \sum_{i} p_{ji} \, \, \, .
\end{equation}
An estimate for the effective transmissibility on a specific day may be obtained by averaging the quantities $R_j$ of Eq.~(\ref{eq:EQ0009}) for all subjects $j$ with onset of symptoms on that day. The original method is 
categorised as a retrospective technique: it yields $R(t)$ estimates at all times $t$ for which all secondary infections have been reported. Consequently, the application of the method yields the effective transmissibility 
with a delay of $M$ (the dimension of the array of weights) days. Modifications, accounting for yet undetected secondary infections, have appeared in the literature, e.g., see Ref.~\cite{Cauchemez2006}.

The diagonal elements of the so-called Wallinga-Teunis upper-triangular matrix vanish: $p_{jj}=0$. Excepting the first subject, the matrix satisfies (by definition): $\sum_{j} p_{ji} = 1$, which implies that only the 
first subject (`patient zero' in the time series) has been `externally' infected. Modifications, accounting for the possibility of external infections, have appeared in the literature, e.g., see Ref.~\cite{Cowling2008}.

A word of caution about the uncertainties $\delta R(t)$. References \cite{Wallinga2004,Obadia2012} recommend the determination of these uncertainties on the basis of simulations. On the contrary, Ref.~\cite{Cowling2008} 
provided an analytical formula (see Appendix therein), but no details about how that formula had been obtained; the starting sentence in their Appendix: ``The methods used to estimate $R_t$ are described in detail elsewhere 
in the literature'' and the subsequent reference to the paper of Wallinga and Teunis apply only to the expectation value of $R(t)$, not to its assigned uncertainty. Considering the authors' second equation, i.e., the 
one which refers to the variance of the $R(t)$ distribution, the origin of the second term within the square brackets on the right-hand side is unclear. The two formulae, given in the Appendix of Ref.~\cite{Cowling2008}, 
found their way into Ref.~\cite{Nishiura2009}, though the copy did not work well for the first of these expressions, the one yielding $R(t)$. In addition, the quantity $q$, entering both expressions (and, according to 
the definitions of Ref.~\cite{Cowling2008}, representing the number of subjects in the time series whose infections did not originate from within that time series), was left undefined in Ref.~\cite{Nishiura2009}.

The introduction of a variant of the Wallinga-Teunis method, accounting for variations in the infectivity of the individuals (e.g., depending on their age group), may be found in Ref.~\cite{Glass2011}.

To summarise, the original Wallinga-Teunis method uses the time series of new cases to extract an estimate for the effective transmissibility; it is a probabilistic method, which employs infectivity weights associated 
with the PDF of the serial interval after discarding its negative tail. Although no attempt will be made herein to include in the method the effects of the pre-symptomatic transmission, the relevant modifications could 
be straightforward.

\subsection{\label{sec:WL}The Wallinga-Lipsitch (WL) method}

Setting forth an analogy between the demographic and the disease-dissemination problems, Wallinga and Lipsitch introduced in 2007 a method for extracting $R(t)$ from the exponential epidemic growth rate $r$ and the PDF 
of the generation interval \cite{Wallinga2007}. To the best of my knowledge, this is the only popular method which may be thought of as resting upon the original definition of the effective reproduction number, which 
(as aforementioned) explicitly takes the recovery interval into account. Within the framework of the SIR model, the authors obtained the approximation: $R(t)=1+r(t)/\gamma$ (applicable for $r(t) > -\gamma$), whereas 
within the SEIR model, they obtained: $R(t)=(1+r(t)/\alpha) \, (1+r(t)/\gamma)$ (applicable for $r(t) > - \min \{ \alpha,\gamma \} $). (The last condition appears as $r(t) > \min \{ -\alpha, -\gamma \}$ in 
Ref.~\cite{Wallinga2007}, which is obviously a mistype.) In the context of the model of this work, it therefore appears that one could use
\begin{equation} \label{eq:EQ0012}
R(t) = (1+r(t)/\alpha) \, (1+r(t)/\avg{\gamma}) \, \, \, ,
\end{equation}
with
\begin{equation} \label{eq:EQ0013}
r(t) = \frac{1}{\tau} \ln \left( \frac{\mathscr{I} (t+\tau; \tau)}{\mathscr{I} (t; \tau)} \right) \, \, \, ,
\end{equation}
where $\mathscr{I} (t; \tau)$ has been defined in Eq.~(\ref{eq:EQ0002}). The appropriate equations for discretised, equidistant (constant time increment $\tau$) data are:
\begin{equation} \label{eq:EQ0014}
R(t_i) = (1 + r_i / \alpha) \, (1 + r_i / \avg{\gamma}) \, \, \, ,
\end{equation}
with
\begin{equation} \label{eq:EQ0015}
r_i = \frac{1}{\tau} \ln \left( \frac{\mathscr{I}_{i+1}}{\mathscr{I}_i} \right) \, \, \, ,
\end{equation}
where $\mathscr{I}_i$ represents in the Wallinga-Lipsitch method the number of \emph{new infections} between the time instants $t_i$ and $t_{i+1}$, rather than the number of subjects with \emph{onset of symptoms} between 
those time instants. To be able to apply the method herein, the two quantities, i.e., the number of new infections on a specific day and the number of subjects with onset of symptoms on that day, must be considered 
identical. As the number of new cases is reported daily in case of the COVID-19 disease, $t_i=i \tau$ for $\tau=1$ d.

The application of the method to low-statistics input is bound to result in sizeable uncertainties. One way to suppress the `noise' is to obtain the growth rates $r_i$ after the application of a filter to the data, e.g., 
featuring a suitable window running over the input time series: in case of linear (two-parameter) fits, a window size of five days (i.e., from element $\mathscr{I}_{i-2}$ to element $\mathscr{I}_{i+2}$) suffices, as does 
one of seven days (similarly centred over the $i$-th datapoint) in case of quadratic (three-parameter) fits.

To summarise, the Wallinga-Lipsitch method uses the time series of new infections to extract an estimate for the effective reproduction number. The authors demonstrate in their paper that the estimate may be obtained as 
the inverse of the value of the moment-generating function of the generation interval at $r=-r_i$, where the exponential epidemic growth rate $r_i$ is given in Eq.~(\ref{eq:EQ0015}); to reduce the fluctuation in the 
results, the determination of the $r_i$ value could involve linear or quadratic fits to the original data. The simple approximation, which the authors obtained within the context of the SEIR model, will be used herein. 
In order that the method be applied in this form, one assumption must be made: that the number of new infections on a specific day is identical to the number of subjects with onset of symptoms on that day. One way to 
avoid this assumption would be to apply the method to the time series of new infections, obtained via a deconvolution from the time series of the new cases after taking the PDFs of the serial and generation intervals 
into account; however, this subject goes beyond the scope of the present work.

\subsection{\label{sec:FNC}The Fraser-Nishiura-Chowell (FNC) method}

The effective transmissibility at time $t$ may be obtained via the relation
\begin{equation} \label{eq:EQ0010}
R(t) = \frac{\mathscr{I} (t; \tau)}{\int_0^t \mathscr{I} (t-x; \tau) w(x) dx} \, \, \, ,
\end{equation}
see Refs.~\cite{Fraser2007,Nishiura2009}. Upon discretisation of this equation, a simple estimator emerges:
\begin{equation} \label{eq:EQ0011}
R(t_i) = \frac{\mathscr{I}_i}{\sum_{j=0}^{M-1} \mathscr{I}_{i-j} w_j} \, \, \, .
\end{equation}

It appears that one could take account of the pre-symptomatic transmission, by extending the infectivity weights into the negative domain; for instance, one could use
\begin{equation} \label{eq:EQ0011_5}
R(t_i) = \frac{\mathscr{I}_i}{\sum_{j=-N}^{M-1} \mathscr{I}_{i-j} w_j} \, \, \, ,
\end{equation}
for weights applying to times between $-N \tau$ and $M \tau$ with respect to the start of the current day. The quantities $N$ and $M$ may be chosen in such a way as to cover most of the region of interest of the PDF of 
the serial interval. However, the exploration of this option lies beyond the scope of the present study.

To summarise, this is a simple approach using the time series of new cases to extract an estimate for the effective transmissibility; it employs infectivity weights associated with the PDF of the serial interval after 
discarding its negative tail. The inclusion in the method of the effects of pre-symptomatic transmission could be straightforward, e.g., see Eq.~(\ref{eq:EQ0011_5}).

\subsection{\label{sec:CFFC}The Cori-Ferguson-Fraser-Cauchemez (CFFC) method}

The essentials of the algorithm of Cori and collaborators may be found in Ref.~\cite{Cori2013}; important details are given in the supplementary material, which is available online. According to that document, the 
authors assume that the rate at which a subject, who was infected on day $i-j<i$, generates new infections (i.e., infects susceptible subjects) on day $i$ may be set equal to the product $R_i w_j$, where $R_i$ is the 
effective transmissibility on day $i$. The probability of observing exactly $\mathscr{I}_i$ new infections on day $i$ is expected to follow the Poisson distribution with parameter $\lambda_i$, where
\begin{equation} \label{eq:EQ0016}
\lambda_i = R_i \Lambda_i \coloneqq R_i \sum_{j=1}^{M-1} \mathscr{I}_{i-j} w_j \, \, \, .
\end{equation}
This probability is equal to
\begin{equation} \label{eq:EQ0017}
P \left( \mathscr{I}_i ; R_i \mid \mathscr{I}_0, \dots , \mathscr{I}_{i-1} ; w \right) = \frac{\lambda_i^{\mathscr{I}_i} e^{-\lambda_i}}{\mathscr{I}_i !} \, \, \, .
\end{equation}
Under the assumption that the effective transmissibility is constant over the last $m>1$ days, one obtains
\begin{equation} \label{eq:EQ0017_1}
P \left( \mathscr{I}_{i-m+1} , \dots , \mathscr{I}_i ; R_{i;m} \mid \mathscr{I}_0, \dots , \mathscr{I}_{i-m} ; w \right) = \prod_{s=i-m+1}^{i} \frac{\lambda_s^{\mathscr{I}_s} e^{-\lambda_s}}{\mathscr{I}_s !} \, \, \, ,
\end{equation}
where obviously
\begin{equation} \label{eq:EQ0017_2}
\lambda_s = R_{i;m} \Lambda_s \, \, \, .
\end{equation}

Starting from a gamma-distributed prior for $R_{i;m}$ with parameters $a$ (shape) and $b$ (scale), the authors obtained the posterior joint distribution
\begin{equation} \label{eq:EQ0017_3}
f(R_{i;m}, \dots) = \frac{1}{\Gamma(a) b^a} R_{i;m}^{a-1} \exp \left( - R_{i;m} / b \right) \prod_{s=i-m+1}^{i} \frac{\lambda_s^{\mathscr{I}_s} e^{-\lambda_s}}{\mathscr{I}_s !} \, \, \, ,
\end{equation}
which implies that
\begin{equation} \label{eq:EQ0017_4}
f(R_{i;m}) \sim R_{i;m}^{\bar{a}-1} \exp \left( - R_{i;m} / \bar{b} \right) \, \, \, ,
\end{equation}
with
\begin{equation} \label{eq:EQ0017_5}
\bar{a} = a + \sum_{s=i-m+1}^i \mathscr{I}_s
\end{equation}
and
\begin{equation} \label{eq:EQ0017_6}
\bar{b}^{-1} = b^{-1} + \sum_{s=i-m+1}^i \Lambda_s \, \, \, .
\end{equation}
Evidently, Eq.~(\ref{eq:EQ0017_4}) suggests that the quantity $R_{i;m}$ follows the gamma distribution with shape parameter $\bar{a}$ given by Eq.~(\ref{eq:EQ0017_5}) and scale parameter $\bar{b}$ given by 
Eq.~(\ref{eq:EQ0017_6}). This results in $\avg{R_{i;m}} = \bar{a} \bar{b}$, whereas the uncertainty $\delta R_{i;m}$ may directly be obtained from the variance of the gamma distribution $\bar{a} \, \bar{b}^2$: 
$\delta R_{i;m} = \sqrt{\bar{a}} \, \bar{b}$.

It is important to note that the weight $w_0$ is dropped in the CFFC method and that the normalisation condition in their case reads as: $\sum_{j=1}^{M-1} w_j = 1$. This implies that the infectors cannot generate 
infections on the day they themselves became infected.

To summarise, there are a number of issues regarding this method. To start with, the same assumption, as in the application of the Wallinga-Lipsitch method, must be made: the number of new infections on a specific day 
is taken to be identical to the number of subjects with onset of symptoms on that day; to circumvent this assumption, one could apply the method to the time series of new infections after deconvolving the input data on 
the basis of the PDFs of the serial and generation intervals. There is no doubt that what is of relevance in this algorithm is the infection-to-infection interval, i.e., the generation interval; the onset of symptoms 
does not enter their approach. In addition, infectors are not capable of generating infections on the day they themselves became infected. It is unclear what weights one would have to use when applying the method 
directly to the time series of new cases, rather than those of new infections. I am not aware of works in which these issues have openly been discussed with a critical eye. For the sake of consistency, the infectivity 
weights, entering the quantities $\Lambda_i$ defined in Eq.~(\ref{eq:EQ0016}), will also be associated herein with the PDF of the serial interval.

\subsection{\label{sec:EffTrCOVID}A short description of selected works addressing the determination of the effective transmissibility in case of the COVID-19 disease}

Estimates for the effective transmissibility in case of the COVID-19 disease have been extracted in a number of works. These studies make use of various PDFs of the serial or of the generation interval, resulting in 
sizeably different average values for these quantities, from about four to over seven days. Some selected studies are listed below.
\begin{itemize}
\item In Ref.~\cite{Shim2020}, the authors made use of the algorithm of Section \ref{sec:FNC} with a modified denominator in Eq.~(\ref{eq:EQ0011}) which could account for infectors who do not appear in the specific time 
series of infections. They then evaluated the effective transmissibility in South Korea as of February 26, 2020, and obtained the result $R=1.5$ for that date, delimited between $1.4$ and $1.6$ with $95~\%$ confidence.
\item Using the CFFC method, $R(t_i)$ was estimated in Ref.~\cite{Pan2020} before and after the introduction of mitigation measures in Wuhan. The authors emphasise the importance of such measures in the attempt to 
harness the dissemination of the COVID-19 disease.
\item In Ref.~\cite{Wang2020}, data from China between January 10 and February 16, 2020, were split into three phases, namely before the `Wuhan lockdown' (January 16 to 22), after `Wuhan lockdown' but before the WHO 
issued a Public Health Emergency of International Concern (PHEIC), and after the PHEIC announcement (January 31 to February 16). Using the WT method, the authors presented results for these three phases for all China, 
as well as (separately) for Hubei and Wuhan. In addition, they discussed the sensitivity of their $R(t_i)$ results to some anomalies in the input, e.g., to the use of underestimated numbers of new cases.
\item In Ref.~\cite{Heiden2020}, Heiden and Hamouda (Robert-Koch Institut) proposed the simple estimator
\begin{equation} \label{eq:EQ0018}
R(t_i) = \frac{\sum_{k=0}^{M-1} \mathscr{I}_{i-M+1+k}}{\sum_{k=0}^{M-1} \mathscr{I}_{i-2M+1+k}} \, \, \, ,
\end{equation}
with $M=4$, hence suggesting the evaluation of $R(t_i)$ as the ratio of the total number of infections within the last two four-day intervals prior to the time instant $t_i$. Their choice of four days relates to the 
expectation value of the average generation interval from recent analyses (which come up with lower estimates when compared to those extracted in the initial phases of the COVID-19 disease). The method will be labelled 
as RKI henceforth. Although the authors mention in their report that $R(t_i)$ of Eq.~(\ref{eq:EQ0018}) should characterise the entire interval of four days ending at time $t_i$, it will be assigned herein to the last 
day ($t_i$). An estimate for the effective transmissibility on any day (after the first full week in each input time series) will thus be obtained on the basis of the number of new cases contained in the eight-day 
interval ending on that day.
\item From data corresponding to the first ten days of the dissemination of the COVID-19 disease in seven countries in Latin America, $R(t_i)$ estimates were obtained following the CFFC method, and were subsequently 
compared with corresponding results for Spain and Italy \cite{Caicedo2020}. The authors warned of a dependence of the $R(t_i)$ estimate on the assumed PDF of the serial/generation interval.
\item Also in Ref.~\cite{Knight2020}, Knight and Mishra discuss the dependence of the $R(t_i)$ estimates on the assumed PDF of the serial/generation interval. Concerning the serial interval, distributions with and 
without negative tails were considered, enabling the authors to conclude that the former PDFs underestimate and the latter overestimate $R(t_i)$.
\item In their report, Gostic and collaborators \cite{Gostic2020} compared $R(t_i)$ results obtained from simulated data using three established methods, concluding that the CFFC method ``most accurately estimates the 
instantaneous reproductive number in real time.'' They addressed the issue of the appropriate PDF in the determination of $R(t_i)$, as well as the effects induced by reporting delays and incomplete observations.
\item Finally, in a long report by the Joint Research Centre (JRC), the European Commission's Science and Knowledge Service, Annunziatio and Asikainen \cite{Annunziatio2020} compared $R(t_i)$ results obtained with 
several methods (some of which have not been mentioned in this work). They also introduced their own method, which may be thought of as a variant of the Bettencourt and Ribeiro method \cite{Bettencourt2008}.
\end{itemize}

In this work, I decided to test the five methods which are listed in Table \ref{tab:MethodsOverview}. As the table reveals, the only method, which explicitly uses input relating to the development of the disease after 
the onset of symptoms, is the one proposed by Wallinga and Lipsitch \cite{Wallinga2007}; the method yields estimates for the effective reproduction number. Evidently, the remaining four methods determine the effective 
transmissibility.

\begin{table}
{\bf \caption{\label{tab:MethodsOverview}}}Some characteristics of the selected algorithms for extracting estimates for the effective reproduction number or for the effective transmissibility in this work.
\vspace{0.2cm}
\begin{center}
\begin{tabular}{|c|c|c|c|c|}
\hline
Method & Means of & Use of details & Use of details & Comments\\
 & extraction & prior to onset & after onset & \\
\hline
\hline
WT \cite{Wallinga2004} & Eqs.~(\ref{eq:EQ0008},\ref{eq:EQ0009}) & Array $w$ linked to & & Delay of $M$ d\\
 & & the serial interval & & \\
\hline
WL \cite{Wallinga2007} & Eq.~(\ref{eq:EQ0014}), linear & $\alpha$ & $\avg{\gamma}$ & Delay of $2$\\
 & or quadratic fits & & & or $3$ d\\
\hline
FNC \cite{Fraser2007,Nishiura2009} & Eq.~(\ref{eq:EQ0011}) & Array $w$ linked to & & \\
 & & the serial interval & & \\
\hline
CFFC \cite{Cori2013} & Bayesian framework, & Array $w$ linked to & & \\
 & Eqs.~(\ref{eq:EQ0017_5},\ref{eq:EQ0017_6}) & the serial interval & & \\
\hline
RKI \cite{Heiden2020} & Eq.~(\ref{eq:EQ0018}) & Constant generation & & \\
 & & interval $\approx \alpha^{-1}$ & & \\
\hline
\end{tabular}
\end{center}
\vspace{0.5cm}
\end{table}

I will pause here and reflect on the consequences in case that all elements of a time series of new cases were scaled (upwards or downwards) by a constant factor: $\mathscr{I}_i \to c \mathscr{I}_i$. The answer to this 
question is that, when using the WL, the FNC, or the RKI method, the estimates for the effective reproduction number or for the effective transmissibility are not affected, whereas small inessential changes are induced 
in the results with the remaining two methods. This observation has one important consequence: provided that the value of the model parameter $p_0$ remains constant with time, the fact that the asymptomatic subjects 
escape detection has no consequences in the evaluation of $R(t)$.

\section{\label{sec:Results}Results}

\subsection{\label{sec:TestResults}Results of the tests}

Using the parameterisation of the temporal dependence of the effective reproduction number of Eq.~(\ref{eq:EQ0004}), time series of new cases were generated via Eq.~(\ref{eq:EQ0002}) for the demographic characteristics 
of the five top-ranking countries in terms of the cumulative number of infections by January 1, 2021, namely the United States of America, India, Brazil, Russia, and the United Kingdom; one subject was assumed to be 
exposed at $t=0$. The fourth-order Runge-Kutta method (see Ref.~\cite{Press2007}, pp.~907--910), with a constant increment of $10$ s, was used in the solution of the system of ODEs of Eqs.~(\ref{eq:EQ0001}) for a temporal 
interval $t_{\rm tot}$ matching the observation interval in the real-life data (as the case is for each country). Some details are given in Table \ref{tab:Characteristics}.

\begin{table}
{\bf \caption{\label{tab:Characteristics}}}Some details about the real-life \cite{ECDC} and the simulated data. The first column corresponds to the country identifier. The adjacent two columns concern the real-life data: 
the first of these columns contains the number of days between the date of report of the first infection(s) and December 14, 2020 (marking the end of the observation interval in Ref.~\cite{ECDC}); listed in the subsequent 
column is the cumulative number of infections by December 14, 2020. The remaining four columns concern the simulated data, as follows. The first of these columns corresponds to the temporal interval $t_{\rm tot}$ within 
which the solution of the system of ODEs of Eqs.~(\ref{eq:EQ0001}) is sought; the second to the identifier of the first day on which the cumulative number of infections exceeds $1$ (subject) in the solution of the system 
of ODEs of Eqs.~(\ref{eq:EQ0001}); the third to the cumulative number of infections at the end of the temporal interval $t_{\rm tot}$; and the fourth to the value of the parameter $P$ of Eq.~(\ref{eq:EQ0004}) achieving 
these results. The demographic characteristics of each country were fixed from Table \ref{tab:CountryRelatedParameters}; in all cases, one subject was assumed to be exposed at $t=0$. The parameter $L$ of 
Eq.~(\ref{eq:EQ0004}) was set equal to $0.90$ in all cases.
\vspace{0.2cm}
\begin{center}
\begin{tabular}{|c||c|c||c|c|c|c|}
\hline
Country & Span (d) & Cum.~number & $t_{\rm tot}$ (d) & First & Cum.~number & $P$\\ 
 & & of infections & & infection (d) & of infections & \\ 
\hline
\hline
USA & $329$ & $16\,256\,754$ & $334$ & $5$ & $16\,255\,469$ & $2.4675$\\
IND & $320$ & $9\,884\,100$ & $325$ & $5$ & $9\,886\,256$ & $2.4485$\\
BRA & $293$ & $6\,901\,952$ & $298$ & $5$ & $6\,903\,642$ & $2.5601$\\
RUS & $318$ & $2\,653\,928$ & $323$ & $5$ & $2\,654\,937$ & $2.3208$\\
GBR & $318$ & $1\,849\,403$ & $323$ & $5$ & $1\,848\,831$ & $2.2843$\\
\hline
\end{tabular}
\end{center}
\vspace{0.5cm}
\end{table}
 
The generated time series of new cases were submitted to the software application aiming at extracting the effective reproduction number or the effective transmissibility on the basis of the algorithms detailed in 
Sections \ref{sec:WT}-\ref{sec:CFFC}, as well as of the RKI method \cite{Heiden2020}. To suppress the noise, the analysis of each time series of new cases commences at the time when the cumulative number of infections 
reaches the (user-defined) threshold of $100$ cases.

The time series of the actual (independent variable) and of the reconstructed (dependent variable) $R_i$ arrays were compared as follows.
\begin{itemize}
\item The two arrays were fitted assuming a linear relation with slope parameter $a$ and intercept parameter $b$. Assuming i) the correctness of each method, ii) the identification of the effective reproduction number 
with the effective transmissibility, and iii) the correctness of the assignment of the resulting $R$ value to a time instant (day), one would expect to obtain from these fits: $a=1$ and $b=0$. A significant departure 
from these expectation values may imply that the particular method is deficient; that the effective reproduction number cannot be identified with the effective transmissibility; or that the assignment of the extracted 
values of the effective reproduction number or of the effective transmissibility to the time is incorrect. In any case however, a departure from the aforementioned expectation values does not entail the inability to put 
forward a suitable correction scheme.
\item Pearson's correlation coefficient $\rho$ was obtained for each pair of actual and reconstructed arrays. In practice (and assuming a linear relationship between the two $R_i$ arrays), the effectiveness of any 
correction scheme increases with increasing $\mid \rho \mid$ value.
\end{itemize}

The results of the comparison between the time series of actual and reconstructed $R_i$ arrays are given in Table \ref{tab:Results}.
\begin{itemize}
\item The first observation is that the results within each method agree reasonably well among themselves for the five countries. Although the results for the different countries will be corrected in this work on the 
basis of the entries in this table, the corrections could also involve weighted averages for $a$ and $b$ for each method.
\item The fitted results for the slope and the intercept, obtained with the WL method \cite{Wallinga2007}, come closer to the expectation values of $1$ and $0$, respectively.
\item Owing to the large $\rho$ values for all methods, the reconstructed results may easily be corrected (on the basis of the aforementioned linear relation) to yield the actual value of the effective reproduction number 
in all cases.
\item Last but not least, all methods systematically underestimate the effective reproduction number when that quantity (i.e., the actual effective reproduction number) exceeds $1$!
\end{itemize}

\begin{table}
{\bf \caption{\label{tab:Results}}}The results of the comparison between the time series of actual and reconstructed $R_i$ arrays. The uncertainties in the slope $a$ and intercept $b$ of the linear fit correspond to one 
standard deviation ($1 \sigma$), and they have been corrected for the quality of each fit via the application of the Birge factor (whenever exceeding $1$).
\vspace{0.2cm}
\begin{center}
\begin{tabular}{|c|c|c|c|}
\hline
\hline
Country & Slope $a$ & Intercept $b$ & Pearson's correlation\\
 & & & coefficient $\rho$\\
\hline
\multicolumn{4}{|c|}{WT method \cite{Wallinga2004}}\\
\hline
USA & $0.4342(16)$ & $0.5293(20)$ & $0.9998$\\
IND & $0.4100(21)$ & $0.5681(26)$ & $1.0000$\\
BRA & $0.4251(24)$ & $0.5405(30)$ & $0.9999$\\
RUS & $0.4162(42)$ & $0.5594(50)$ & $1.0000$\\
GBR & $0.4215(51)$ & $0.5516(61)$ & $0.9999$\\
\hline
\multicolumn{4}{|c|}{WL method \cite{Wallinga2007}, linear fits}\\
\hline
USA & $0.93288(55)$ & $-0.00248(76)$ & $0.9961$\\
IND & $0.89070(45)$ & $0.06798(60)$ & $0.9951$\\
BRA & $0.92039(50)$ & $0.01414(69)$ & $0.9971$\\
RUS & $0.89794(52)$ & $0.05619(66)$ & $0.9950$\\
GBR & $0.90729(60)$ & $0.04177(74)$ & $0.9880$\\
\hline
\multicolumn{4}{|c|}{FNC method \cite{Fraser2007,Nishiura2009}}\\
\hline
USA & $0.4148(14)$ & $0.5610(17)$ & $0.9993$\\
IND & $0.3924(18)$ & $0.5967(21)$ & $0.9995$\\
BRA & $0.4063(20)$ & $0.5731(24)$ & $0.9994$\\
RUS & $0.3981(37)$ & $0.5880(42)$ & $0.9992$\\
GBR & $0.4030(44)$ & $0.5806(51)$ & $0.9987$\\
\hline
\end{tabular}
\end{center}
\vspace{0.5cm}
\end{table}

\begin{table*}
{\bf Table \ref{tab:Results} continued}
\vspace{0.2cm}
\begin{center}
\begin{tabular}{|c|c|c|c|}
\hline
\hline
Country & Slope $a$ & Intercept $b$ & Pearson's correlation\\
 & & & coefficient $\rho$\\
\hline
\multicolumn{4}{|c|}{CFFC method \cite{Cori2013}, $m=7$ d}\\
\hline
USA & $0.45283(58)$ & $0.52710(67)$ & $0.9999$\\
IND & $0.42859(75)$ & $0.56560(86)$ & $1.0000$\\
BRA & $0.44410(84)$ & $0.54080(98)$ & $0.9999$\\
RUS & $0.4346(15)$ & $0.5560(17)$ & $1.0000$\\
GBR & $0.4399(18)$ & $0.5479(21)$ & $1.0000$\\
\hline
\multicolumn{4}{|c|}{RKI method \cite{Heiden2020}}\\
\hline
USA & $0.3233(11)$ & $0.6575(12)$ & $0.9996$\\
IND & $0.3064(14)$ & $0.6847(16)$ & $0.9994$\\
BRA & $0.3177(15)$ & $0.6658(18)$ & $0.9994$\\
RUS & $0.3104(27)$ & $0.6785(31)$ & $0.9992$\\
GBR & $0.3140(33)$ & $0.6729(38)$ & $0.9985$\\
\hline
\end{tabular}
\end{center}
\vspace{0.5cm}
\end{table*}

\subsection{\label{sec:Application}Extraction of the effective transmissibility in the real-life data}

As mentioned in Section \ref{sec:Introduction}, the corrections extracted in Section \ref{sec:TestResults} will be applied to the real-life time series of new cases obtained from the European Centre for Disease Prevention 
and Control \cite{ECDC}. With regard to the (common) start of the observation interval ($t=0$ at 00:00:00 UTC on December 31, 2019), the threshold of $100$ cumulative reported infections was reached on different dates 
for the five countries:
\begin{itemize}
\item USA: on March 3, 2020 (day identifier $i=63$);
\item IND: on March 17, 2020 (day identifier $i=77$);
\item BRA: on March 15, 2020 (day identifier $i=75$);
\item RUS: on March 18, 2020 (day identifier $i=78$); and
\item GBR: on March 4, 2020 (day identifier $i=64$).
\end{itemize}

\begin{figure}
\begin{center}
\includegraphics [width=12cm] {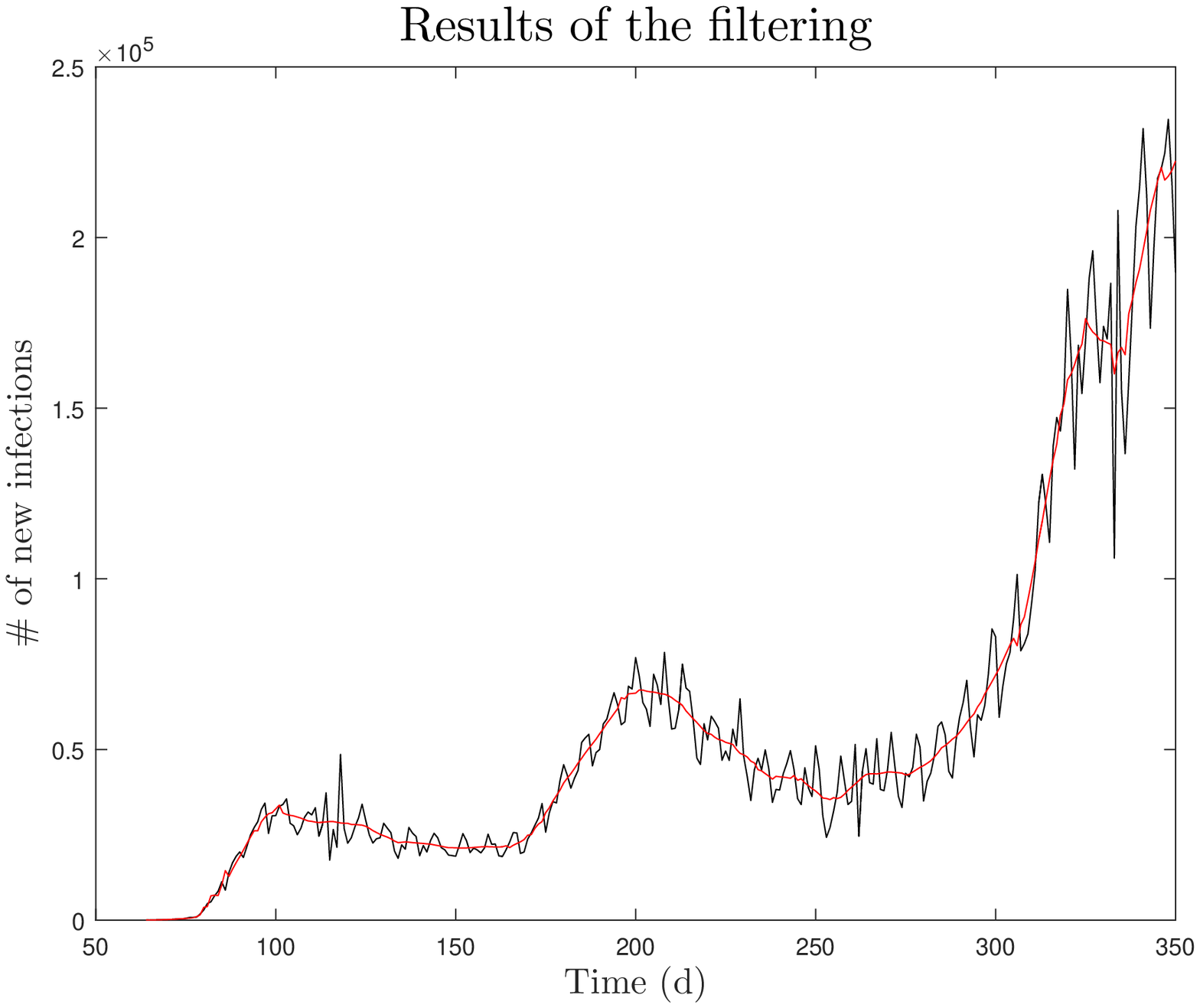}
\caption{\label{fig:FilteredDataUSA}The original and the filtered \cite{Cleveland1988} time series of new cases for the USA. The original data have been obtained from the European Centre for Disease Prevention and 
Control \cite{ECDC}; the observations start on December 31, 2019 and end on December 14, 2020. The filtered data have been obtained via the application of the robust, locally weighted regression algorithm known as LOWESS 
\cite{Cleveland1988}.}
\vspace{0.35cm}
\end{center}
\end{figure}

To suppress the noise which is present in the original data, the CDFs, obtained from the five real-life time series, were filtered by means of the robust, locally weighted regression algorithm known as LOWESS 
\cite{Cleveland1988}; linear fits were performed within each position of the running window, and the fitted value for the central datapoint replaced the original entry in the time series. For each of the five countries, 
a time series of new cases was obtained from the filtered CDF data. As a result, several artefacts, which are present in the original data (e.g., undulating behaviour, enigmatic peaks, etc.), were nearly removed, e.g., 
see Fig.~\ref{fig:FilteredDataUSA} for a comparison between the original and the filtered time series of new cases for the USA.

\begin{figure}
\begin{center}
\includegraphics [width=12cm] {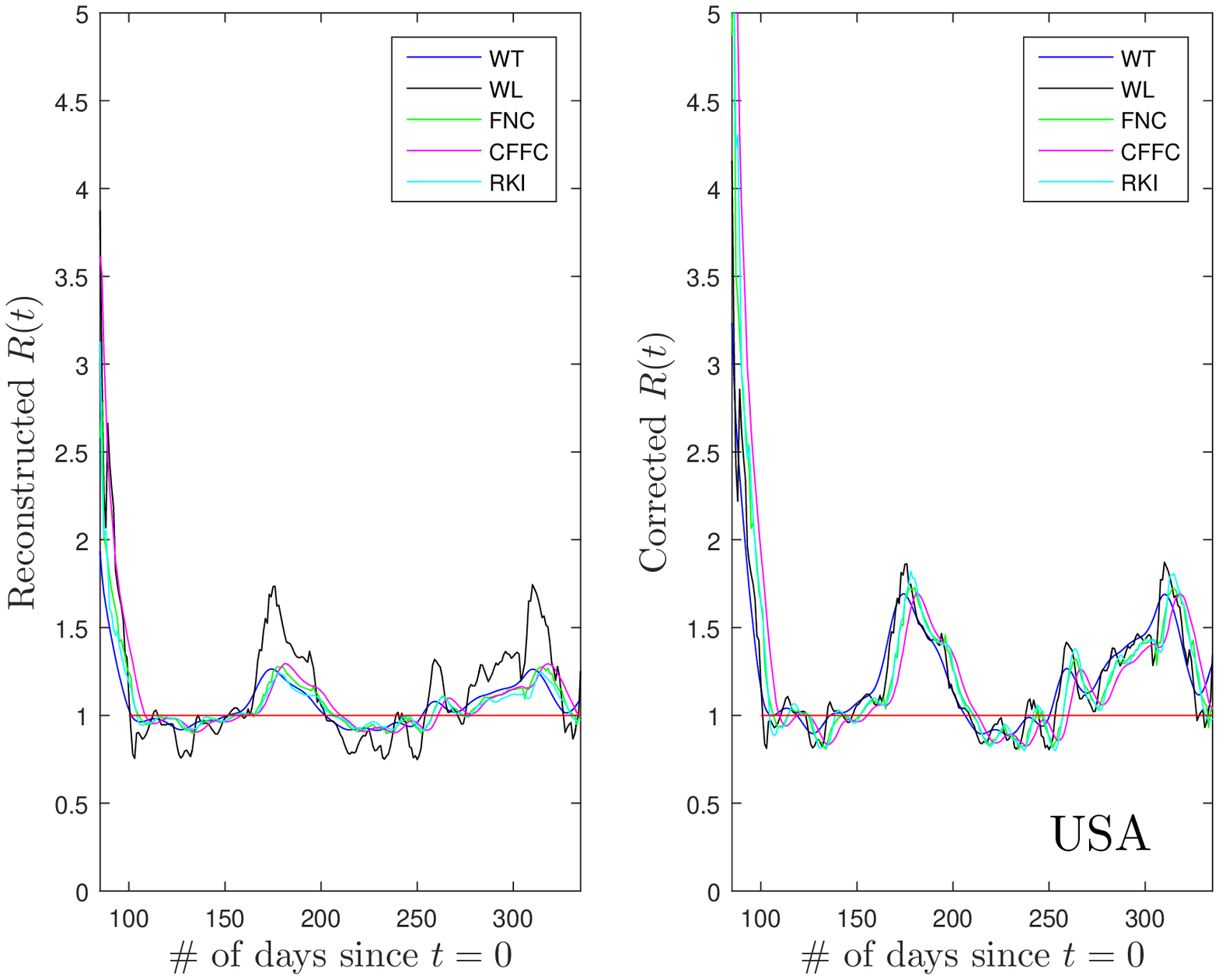}
\caption{\label{fig:ReffUSA}The time series of the reconstructed effective reproduction number or of the reconstructed effective transmissibility for the United States of America, alongside the corrected effective 
reproduction number, obtained from the reconstructed data on the basis of the corrections of Table \ref{tab:Results}. The data analysis starts on March 3, 2020 (day $63$ from the start of the observation interval), when 
the threshold of $100$ cumulative infections was reached. In addition, a short transition phase in each time series has been removed. Two hump-shaped protrusions are clearly visible in the corrected data: one between 
the beginning of June and mid July, the other from the beginning of September to the end of the observation interval.}
\vspace{0.35cm}
\end{center}
\end{figure}

\begin{figure}
\begin{center}
\includegraphics [width=12cm] {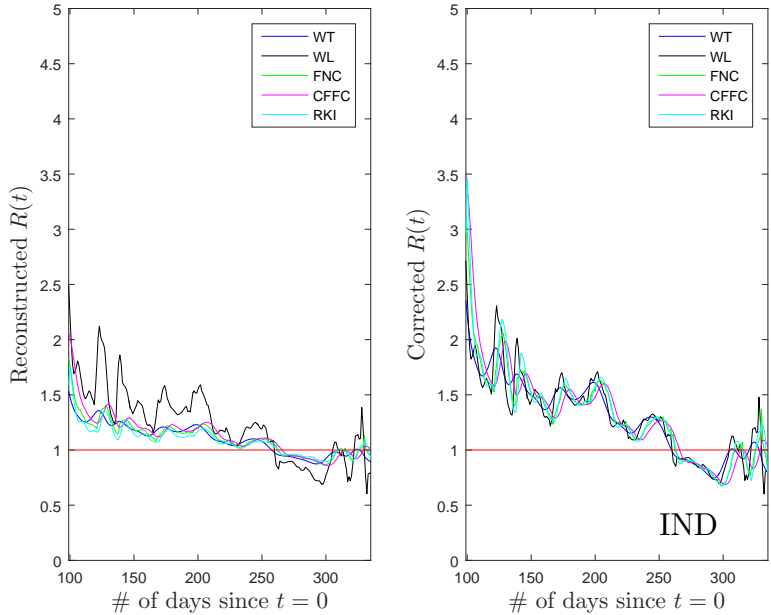}
\caption{\label{fig:ReffIND}Same as Fig.~\ref{fig:ReffUSA} for India. The data analysis starts on March 17, 2020 (day $77$ from the start of the observation interval). The effective reproduction number slowly decreased 
throughout 2020, but remained above the critical value of $1$ until the beginning of September, then undulated around $1$ from the beginning of November to the end of the observation interval.}
\vspace{0.35cm}
\end{center}
\end{figure}

\begin{figure}
\begin{center}
\includegraphics [width=12cm] {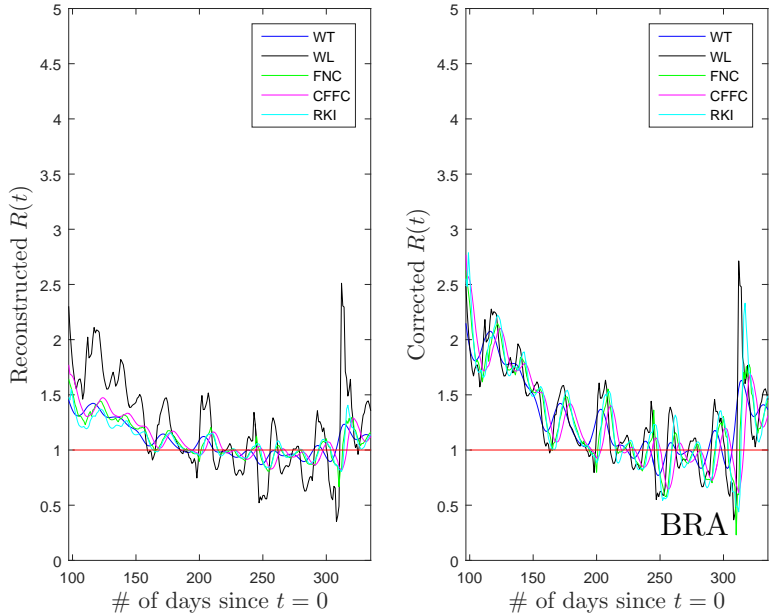}
\caption{\label{fig:ReffBRA}Same as Fig.~\ref{fig:ReffUSA} for Brazil. The data analysis starts on March 15, 2020 (day $75$ from the start of the observation interval). A strongly undulating behaviour is clearly visible 
in the data. The effective reconstruction number exceeded $1$ until about mid July, then widely fluctuated around that value, before returning to values above $1$ from mid October to the end of the observation interval.}
\vspace{0.35cm}
\end{center}
\end{figure}

\begin{figure}
\begin{center}
\includegraphics [width=12cm] {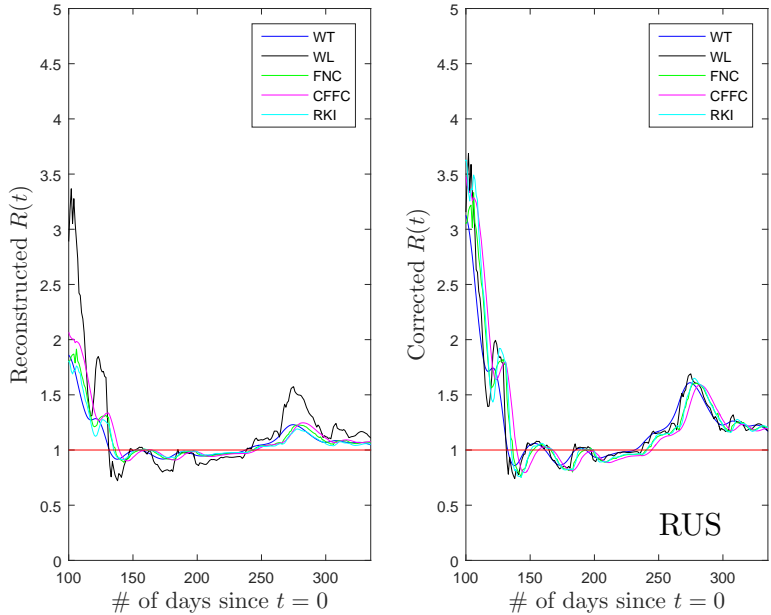}
\caption{\label{fig:ReffRUS}Same as Fig.~\ref{fig:ReffUSA} for Russia. The data analysis starts on March 18, 2020 (day $78$ from the start of the observation interval). The effective reconstruction number dropped to 
values around $1$ by mid May, where it remained until the end of August; around that time however, it bounced back to values exceeding $1$, which persisted until the end of the observation interval.}
\vspace{0.35cm}
\end{center}
\end{figure}

\begin{figure}
\begin{center}
\includegraphics [width=12cm] {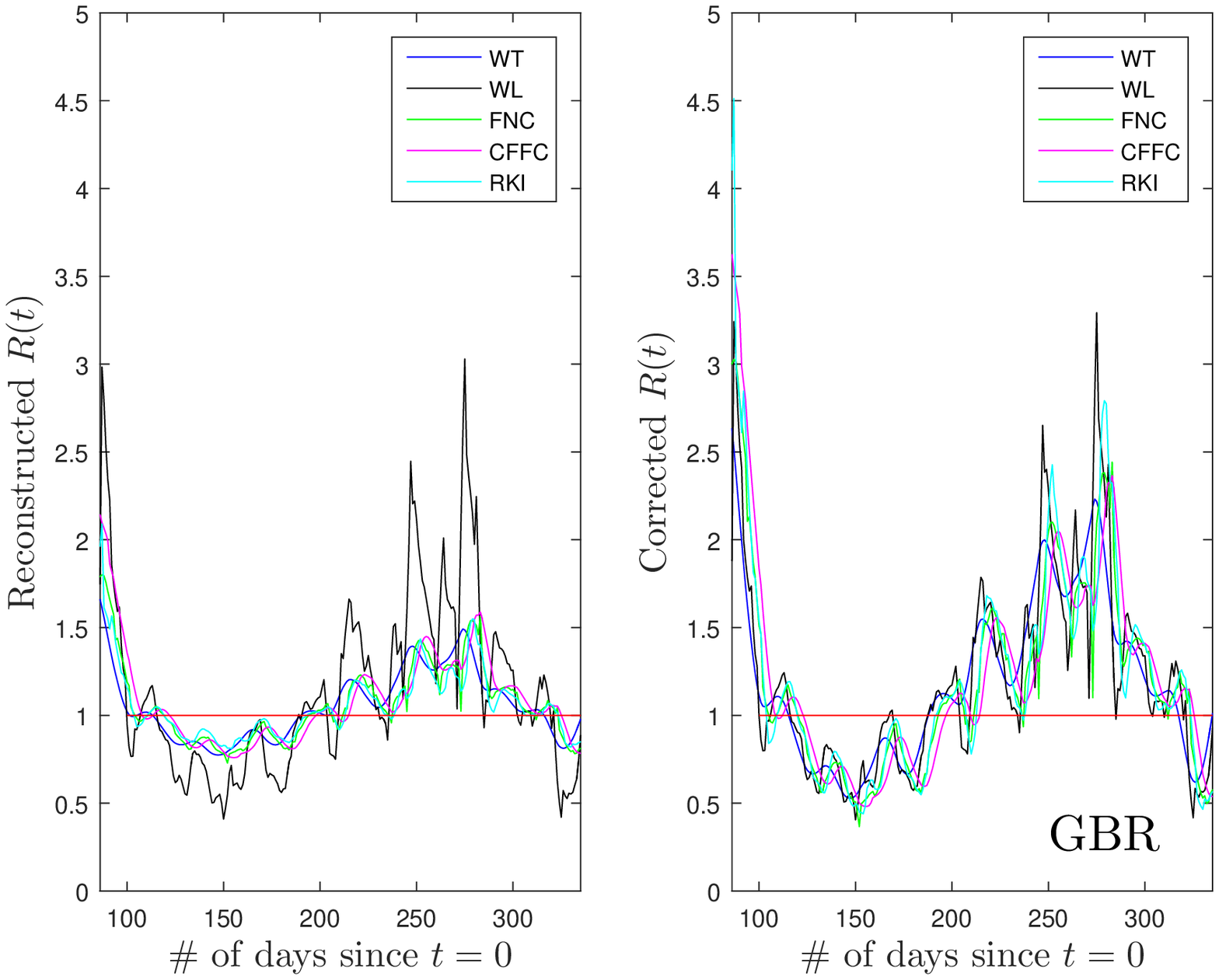}
\caption{\label{fig:ReffGBR}Same as Fig.~\ref{fig:ReffUSA} for the United Kingdom. The data analysis starts on March 4, 2020 (day $64$ from the start of the observation interval). The effective reconstruction number 
undulated around $1$ for most of the observation interval. A sizeable background `bump' is seen in the corrected values for about three and a half months starting from the beginning of August.}
\vspace{0.35cm}
\end{center}
\end{figure}

The extraction of the reconstructed effective reproduction number or of the effective transmissibility was subsequently pursued using the methods of Table \ref{tab:MethodsOverview}, and the results were corrected via 
the application of the results of Table \ref{tab:Results}; the uncertainties, quoted in that table, were not used. The time series of the reconstructed effective reproduction number or of the effective transmissibility, 
as well as the corrected effective reproduction number, obtained from the reconstructed data, are shown in Figs.~\ref{fig:ReffUSA}-\ref{fig:ReffGBR}. In spite of the common (in this analysis) infectivity weights, there 
are differences in the results of the methods which aim at determining the effective transmissibility from the time series of new cases, including shifts in the time domain, which question the assignment of the result of 
(at least some of) these methods to a time.

It is evident that, though four of the methods suggest that the effective transmissibility was nearing $1$ in a large part of the observation interval, the effective reproduction number remained significantly larger. 
Only the reconstructed $R(t)$, extracted with the WL method, comes close to the actual value of the effective reproduction number; however, experience shows that that method is not used as frequently as the methods 
aiming at extracting the effective transmissibility, in particular as frequently as the WT and CFFC methods.

\section{\label{sec:Conclusions}Conclusions and discussion}

This work distinguishes between the effective reproduction number and the effective transmissibility.
\begin{itemize}
\item The former is defined as the ratio of the infection and the recovery rates, and represents the proper indicator when it comes down to assessing the short-term effects and planning the tactics during the course of 
the dissemination of infectious diseases such as the COVID-19 disease, in particular in terms of the hospital preparedness and the demands for intensive-care and ventilation beds.
\item The latter is defined as the average number of infections which an infected subject will generate among his/her contacts, i.e., the average number of secondary infections per infected subject.
\end{itemize}
Both quantities may be evaluated on the basis of analyses of the time series of new cases, and both make use of some characteristics of the disease prior to the onset of symptoms, namely of the probability density 
function (PDF) of the serial interval, in its discretised form. In addition, the methods aiming at determining the effective reproduction number take explicit account of some characteristics of the development of the 
disease after the onset of symptoms, e.g., relating to the recovery interval. For the methods aiming at determining the effective transmissibility, the recovery interval is of no relevance.

The majority of the established algorithms for assessing whether the dissemination of a disease is waxing or waning provide estimates for the effective transmissibility; four such methods have been tested in this work: 
the probabilistic method of Wallinga and Teunis \cite{Wallinga2004}, the method of Fraser, Nishiura, and Chowell \cite{Fraser2007,Nishiura2009} which is based on the renewal equation, the Bayesian framework of Cori and 
collaborators \cite{Cori2013}, and a method developed at the Robert-Koch Institut by Heiden and Hamouda \cite{Heiden2020}. I am aware of just one method which outputs the effective reproduction number, namely the method 
of Wallinga and Lipsitch \cite{Wallinga2007}.

This work establishes a relation between the results of the aforementioned methods and the actual value of the effective reproduction number, using generated data of new cases. These data were obtained on the basis of a 
deterministic compartmental epidemiological model with seven groups, tailored to the characteristics of the COVID-19 disease \cite{Matsinos2020a}; each generated time series of new cases has been obtained with 
Eq.~(\ref{eq:EQ0002}) from the solution of the system of first-order ordinary differential equations (ODEs) of Eqs.~(\ref{eq:EQ0001}). As the objective herein was the application of corrections to real-life time series 
of new cases, the demographic characteristics of the five countries, for which the corrections would apply, were taken into account in the generated data of new cases. Chosen to be the five top-ranking countries in 
terms of the cumulative number of infections by January 1, 2021, these countries are (in descending order of infections): the United States of America, India, Brazil, Russia, and the United Kingdom.

The values of the model parameters, corresponding to the generated data, are detailed in Table \ref{tab:DiseaseRelatedParameters}; the quoted uncertainties were not used. The analysis of each generated time series of 
new cases led to the extraction of corrections between the output of each of the aforementioned methods and the known effective reproduction number used as input in the generation of the time series of new cases: the 
relevant corrections are detailed in Table \ref{tab:Results}. The results suggest that all the aforementioned methods underestimate the effective reproduction number (by variable amounts) when that quantity (i.e., the 
actual effective reproduction number) exceeds $1$. As the quantity, determined in the Wallinga-Lipsitch method, is the effective reproduction number (and not the effective transmissibility), it might not be surprising 
that the results, obtained with this method, come closer to the actual value of the effective reproduction number.

The corrections of Table \ref{tab:Results} were applied to the results of the analysis of five real-life time series of new cases. Figures \ref{fig:ReffUSA}-\ref{fig:ReffGBR} demonstrate that the methods, determining the 
effective transmissibility, underestimate the effective reproduction number. This observation might explain why the reports in the media in 2020 of dropping effective transmissibility in several sovereign countries, even 
below the critical value of $1$ in some cases, were hardly accompanied by a substantial relief of the pressure on the healthcare systems~\footnote{The analysis of the daily reported data on the demands for ordinary 
hospital beds in $29$ European countries \cite{hospitalisationEU} since the beginning of the pandemia reveal that only between April and August 2020 was that demand decreasing with time, from an average of about $340$ 
beds per million subjects at the end of April to less than $50$ at the end of August. In spite of the mitigation measures in most of these countries, the demands remain high since the beginning of November 2020, 
undulating between $300$ and $450$ beds. The same behaviour is clearly seen in the demand for ICU beds, which also remains high since the beginning of November 2020, undulating between $40$ and $55$ beds per million 
subjects. In both cases, the aforementioned results represent averages for the entirety of the population within the $29$ countries. Owing to a number of reasons, there is significant variation in the corresponding results 
across these countries. Regarding this part of the analysis, the details are available upon request.}. The bottom line is that, provided that the recovery interval of an infectious disease is sufficiently longer than the 
time scale associated with the infection rate, even a low value of the effective transmissibility, even below $1$, is bound to generate (in the short term) a backlog of subjects needing hospitalisation.

My final comment is that the effective transmissibility is a misleading indicator of the short-term development of the disease and of the healthcare demands emerging thereof. I believe that the sham impression which this 
quantity generates might have even caused an unintentional relaxation of the mitigation measures during 2020. I cannot explain why the effective transmissibility is so broadly used, inasmuch as the procedure to obtain 
estimates for the appropriate quantity, i.e., for the effective reproduction number, is equally straightforward and less nebulous.

One word of caution. The overall agreement between the corrected results of the effective reproduction number in Figs.~\ref{fig:ReffUSA}-\ref{fig:ReffGBR} among the five methods should not be taken as suggestive of the 
correctness of the main conclusion of this work; it rather attests to its self-consistency. A number of issues deserve further study.
\begin{itemize}
\item A procedure must be developed in order that the PDFs of the quantities entering the system of ODEs of Eqs.~(\ref{eq:EQ0001}), rather than the average values of these distributions (model parameters: $\alpha$, 
$\avg{\gamma}$, $\gamma_1$, $\gamma_2$), be taken into account. In case of a Monte-Carlo approach, this would have been a straightforward subject to deal with; it is less straightforward to provide a solution, while 
retaining the deterministic time evolution of the approach based on the system of ODEs of Eqs.~(\ref{eq:EQ0001}).
\item The identification of the infectivity interval with the recovery interval in the model of this work (see end of Section \ref{sec:Model_Parameters}) must be re-examined. In addition, the hypothesis that the 
transmission probability of the disease is identical for the infective subjects of the three groups $I_{0,1,2}$ must be tested.
\item The probabilities $p_{0, \dots , 3}$ must be re-assessed, taking also account of any recent results. Ways must be investigated to also include in the analysis the uncertainties of the various constants which enter 
the model of this work.
\item The variability of all input parameters, due to the characteristics of the different virus variants, must be studied.
\end{itemize}
In fact, most of the issues above would have been facilitated, had a centralised database of COVID-19 cases been available (see next paragraph). The importance of the exploration of all aforementioned issues notwithstanding, 
I doubt that such effects can have a sizeable bearing on the main conclusion of this work, namely that the methods, which have been tested herein, tend to underestimate the effective reproduction number when that quantity 
(i.e., the actual effective reproduction number) exceeds $1$.

Although not directly related to this work, there is one idea which I would like to share with others. My belief is that a centralised database of COVID-19 cases will be very useful as far as the process of deciphering 
the details of this virus is concerned. For the sake of example, I find it deplorable that the results of several studies, reporting on the PDFs of the incubation/serial/generation interval, are so strikingly different~\footnote{I 
mention in passing two results from studies reporting on the incubation interval: $4.00(11)$ and $8.00(37)$ d; and two from studies reporting on the serial interval: $4.22(40)$ and $7.00(59)$ d.}. In my judgement, the 
first step forwards involves the creation of a reliable database of clear-cut cases by an independent authority in the Medical Domain, e.g., by an organisation such as WHO or by an institution with extensive experience 
in medical procedures, e.g., by the School of Medicine at the Johns Hopkins University; non-profit organisations could also take part in this effort. The work may appear to be overwhelming, yet I believe that it is necessary.

One last comment is due. Several of the studies, aiming at extracting estimates for the incubation/serial/generation interval, do not report any details about the cases their results are based on. Had I been involved in 
the peer-reviewing process, I would not have recommended such papers for publication. Although some readers might find this point of view extreme, I believe that it is justifiable practice when public health comes into 
play. I must mention that, for the purposes of an earlier study \cite{Matsinos2020b}, I had requested the data on the incubation interval from all the works which had earlier reported on that quantity; my intention 
was to combine the distributions and enhance statistics. Unfortunately, half of the corresponding authors did not bother to communicate even a negative decision. In stark contrast, the team of the `Oxford Martin 
Programme on Global Development' at the University of Oxford declares: ``Our work belongs to everyone'' \cite{ourworldindata}; I find their attitude commendable.

\begin{ack}
I would like to thank Jesse Knight and Christophe Fraser for answering my questions regarding Refs.~\cite{Knight2020} and \cite{Cori2013}, respectively.

Figure \ref{fig:CompartmentalModelForCOVID19} has been created with CaRMetal \cite{CaRMetal}. The remaining figures have been created with MATLAB$^{\textregistered}$ (The MathWorks, Inc., Natick, Massachusetts, United 
States).

This study received no funding. I have no affiliations with or involvement in any organisation, institution, company, or firm with financial interest in the subject matter of this work.
\end{ack}


\begin{thebibliography}{99}
\bibitem{Matsinos2020a} E.~Matsinos, `A compartmental epidemiological model for the dissemination of the COVID-19 disease', Preprints 2020060039 (2020). DOI: 10.20944/preprints202006.0039.v2
\bibitem{fda} https://www.fda.gov/emergency-preparedness-and-response/coronavirus-disease-2019-covid-19/covid-19-vaccines
\bibitem{ema} https://www.ema.europa.eu/en/human-regulatory/overview/public-health-threats/coronavirus-disease-covid-19/treatments-vaccines/covid-19-vaccines
\bibitem{ProgressVaccination} https://ourworldindata.org/covid-vaccinations
\bibitem{ECDC} https://www.ecdc.europa.eu/en/covid-19/data
\bibitem{Hethcote2000} H.W.~Hethcote, `The Mathematics of infectious diseases', SIAM Rev.~42, 599 (2000); available from https://www.jstor.org/stable/2653135
\bibitem{Gudbjartsson2020} D.F.~Gudbjartsson, `Humoral immune response to SARS-CoV-2 in Iceland', N.~Engl.~J.~Med.~383, 1724 (2020). DOI: 10.1056/NEJMoa2026116
\bibitem{Ripperger2020} T.J.~Ripperger \etal, `Orthogonal SARS-CoV-2 serological assays enable surveillance of low-prevalence communities and reveal durable humoral immunity', Immunity 53, 925 (2020). DOI: 
10.1016/j.immuni.2020.10.004
\bibitem{Rodda2021} L.B.~Rodda \etal, `Functional SARS-CoV-2-specific immune memory persists after mild COVID-19', Cell 184, 169 (2021). DOI: 10.1016/j.cell.2020.11.029
\bibitem{Ganyani2020} T.~Ganyani \etal, `Estimating the generation interval for coronavirus disease (COVID-19) based on symptom onset data, March 2020', Euro Surveill.~25(17), 2000257 (2020). DOI: 10.2807/1560-7917.ES.2020.25.17.2000257
\bibitem{Wei2020} W.E.~Wei \etal, `Presymptomatic Transmission of SARS-CoV-2 - Singapore, January 23-March 16, 2020', MMWR Morb.~Mortal.~Wkly.~Rep.~69(14), 411 (2020). DOI: 10.15585/mmwr.mm6914e1
\bibitem{Arons2020} M.M.~Arons \etal, `Presymptomatic SARS-CoV-2 infections and transmission in a skilled nursing facility', N.~Engl.~J.~Med.~382(22), 2081 (2020). DOI: 10.1056/NEJMoa2008457
\bibitem{Ren2021} X.~Ren \etal, `Evidence for pre-symptomatic transmission of coronavirus disease 2019 (COVID-19) in China', Influenza Other Respir.~Viruses 15, 19 (2021). DOI: 10.1111/irv.12787
\bibitem{Zhang2021} Y.~Zhang \etal, `Role of presymptomatic transmission of COVID-19: evidence from Beijing, China', J.~Epidemiol.~Community Health 75, 84 (2021). DOI: 10.1136/jech-2020-214635
\bibitem{Caicedo2020} Y.~Caicedo-Ochoa, D.E.~Rebell\'on-S\'anchez, M.~Pe\~naloza-Rall\'on, H.F.~Cort\'es-Motta, Y.R.~M\'endez-Fandi\~no, `Effective reproductive number estimation for initial stage of COVID-19 pandemic 
in Latin American Countries', Int.~J.~Infect.~Dis.~ 95, 316 (2020). DOI: 10.1016/j.ijid.2020.04.069
\bibitem{Knight2020} J.~Knight, S.~Mishra, `Estimating effective reproduction number using generation time versus serial interval, with application to COVID-19 in the Greater Toronto Area, Canada', Infect.~Dis.~Model.~5, 
889 (2020). DOI: 10.1016/j.idm.2020.10.009
\bibitem{Knight2021} J.~Knight, private communication.
\bibitem{PRB} https://www.prb.org/international/indicator/births/table
\bibitem{CIA} https://web.archive.org/web/20201010211308/\\https://www.cia.gov/library/publications/the-world-factbook/fields/346.html
\bibitem{worldometers} https://www.worldometers.info/coronavirus
\bibitem{WP} https://worldpopulationreview.com/countries
\bibitem{Lipsitch2003} M.~Lipsitch \etal, `Transmission dynamics and control of Severe Acute Respiratory Syndrome', Science 300, 1966 (2003). DOI: 10.1126/science.1086616
\bibitem{Cauchemez2006} S.~Cauchemez \etal, `Real-time estimates in early detection of SARS', Emerg.~Infect.~Dis.~12(1), 110 (2006). DOI: 10.3201/eid1201.050593
\bibitem{Farrington2003} C.P.~Farrington, H.J.~Whitaker, `Estimation of effective reproduction numbers for infectious diseases using serological survey data', Biostatistics 4, 621 (2003). DOI: 10.1093/biostatistics/4.4.621
\bibitem{Bettencourt2008} L.M.A.~Bettencourt, R.M.~Ribeiro, `Real time Bayesian estimation of the epidemic potential of emerging infectious diseases', PLoS ONE 3, e2185 (2008). DOI: 10.1371/journal.pone.0002185
\bibitem{Cintron2009} A.~Cintr{\'o}n-Arias, C.~Castillo-Ch{\'a}vez, L.M.A.~Bettencourt, A.L.~Lloyd, H.T.~Banks, `The estimation of the effective reproductive number from disease outbreak data', Math.~Biosci.~Eng.~6, 261 
(2009). DOI: 10.3934/mbe.2009.6.261
\bibitem{Nishiura2009} H.~Nishiura, G.~Chowell, `The effective reproduction number as a prelude to statistical estimation of time-dependent epidemic trends', \emph{Mathematical and Statistical Estimation Approaches in 
Epidemiology}, eds.~G.~Chowell, J.M.~Hyman, L.M.A.~Bettencourt, C.~Castillo-Ch{\'a}vez, Springer, Dordrecht (2009). DOI: 10.1007/978-90-481-2313-1\_5
\bibitem{Wallinga2004} J.~Wallinga, P.~Teunis, `Different epidemic curves for Severe Acute Respiratory Syndrome reveal similar impacts of control measures', Am.~J.~Epidemiol.~160, 509 (2004). DOI: 10.1093/aje/kwh255
\bibitem{Cowling2008} B.J.~Cowling, L.M.~Ho, G.M.~Leung, `Effectiveness of control measures during the SARS epidemic in Beijing: a comparison of the $R_t$ curve and the epidemic curve', Epidemiol.~Infect.~136, 562 (2008). 
DOI: 10.1017/S0950268807008722
\bibitem{Obadia2012} T.~Obadia, R.~Haneef, P.-Y.~Bo{\"e}lle, `The R0 package: a toolbox to estimate reproduction numbers for epidemic outbreaks', BMC Med.~Inform.~Decis.~Mak.~12:147 (2012). DOI: 10.1186/1472-6947-12-147
\bibitem{Glass2011} K.~Glass, G.N.~Mercer, H.~Nishiura, E.S.~McBryde, N.G.~Becker, `Estimating reproduction numbers for adults and children from case data', J.~R.~Soc.~Interface 8, 1248 (2011). DOI: 10.1098/rsif.2010.0679
\bibitem{Wallinga2007} J.~Wallinga, M.~Lipsitch, `How generation intervals shape the relationship between growth rates and reproductive numbers', Proc.~R.~Soc.~B 274, 599 (2007). DOI: 10.1098/rspb.2006.3754
\bibitem{Fraser2007} C.~Fraser, `Estimating individual and household reproduction numbers in an emerging epidemic', PLoS ONE 2(8), e758 (2007). DOI: 10.1371/journal.pone.0000758
\bibitem{Cori2013} A.~Cori, N.M.~Ferguson, C.~Fraser, S.~Cauchemez, `A new framework and software to estimate time-varying reproduction numbers during epidemics', Am.~J.~Epidemiol.~178, 1505 (2013). DOI: 10.1093/aje/kwt133
\bibitem{Shim2020} E.~Shim, A.~Tariq, W.~Choi, Y.~Lee, G.~Chowell, `Transmission potential and severity of COVID-19 in South Korea', Int.~J.~Infect.~Dis.~ 93, 339 (2020). DOI: 10.1016/j.ijid.2020.03.031
\bibitem{Pan2020} A.~Pan \etal, `Association of public health interventions with the epidemiology of the COVID-19 outbreak in Wuhan, China', JAMA 323(19), 1915 (2020). DOI: 10.1001/jama.2020.6130
\bibitem{Wang2020} K.~Wang \etal, `Real-time estimation of the reproduction number of the novel coronavirus disease (COVID-19) in China in 2020 based on incidence data', Ann.~Transl.~Med.~8(11), 689 (2020). DOI: 
10.21037/atm-20-1944
\bibitem{Heiden2020} M.~Heiden, Q.~Hamouda, `Sch\"atzung der aktuellen Entwicklung der SARS-CoV-2-Epidemie in Deutschland - Nowcasting', Epid.~Bull.~17, 10 (2020). DOI: 10.25646/6692.4
\bibitem{Gostic2020} K.M.~Gostic \etal, `Practical considerations for measuring the effective reproductive number, $R_t$', PLoS Comput.~Biol.~16(12), e1008409 (2020). DOI: 10.1371/journal.pcbi.1008409
\bibitem{Annunziatio2020} A.~Annunziatio, T.~Asikainen, `Effective reproduction number estimation from data series', EUR 30300 EN, Publications Office of the European Union, Luxembourg (2020). ISBN: 9789276207498. 
DOI: 10.2760/036156, JRC121343
\bibitem{Press2007} W.H.~Press, S.A.~Teukolsky, W.T.~Vetterling, B.P.~Flannery, `Numerical Recipes: The Art of Scientific Computing' (3rd edn.), Cambridge University Press (2007). ISBN: 9780521880688.
\bibitem{Cleveland1988} W.S.~Cleveland, S.J.~Devlin, `Locally Weighted Regression: an approach to regression analysis by local fitting', J.~Am.~Stat.~Assoc.~83(403), 596 (1988). DOI: 10.1080/01621459.1988.10478639
\bibitem{hospitalisationEU} https://www.ecdc.europa.eu/en/publications-data/download-data-hospital-and-icu-admission-rates-and-current-occupancy-covid-19; https://coronavirus.data.gov.uk/details/healthcare; 
https://en.wikipedia.org/wiki/COVID-19\_pandemic\_in\_Greece
\bibitem{Matsinos2020b} E.~Matsinos, `COVID-19: On the quarantine duration after short visits to high-risk regions', {\tt arXiv:2010.02688 [physics.soc-ph]}.
\bibitem{ourworldindata} https://ourworldindata.org
\bibitem{CaRMetal} https://carmetal.en.uptodown.com/windows
\end{thebibliography}
\end{document}